\documentclass{jfm}

\usepackage{graphicx}
\usepackage{newtxtext}
\usepackage{newtxmath}
\usepackage{natbib}  
\usepackage{hyperref}
\usepackage[normalem]{ulem}
\hypersetup{
    colorlinks = true,
    urlcolor   = blue,
    citecolor  = black,
}

\newcommand{\RomanNumeralCaps}[1]
 \linenumbers

\newcommand{\firstref}[1]{#1} 
\newcommand{\secondref}[1]{#1} 
\newcommand{\thirdref}[1]{#1} 
\newcommand{\fourthref}[1]{#1} 

\newcommand{\Topt}{\mathcal{T}}
\renewcommand{\vec}[1]{\boldsymbol{#1}}

\title{The minimal seed for transition to convective turbulence in heated pipe flow}
\author{Shijun Chu\aff{1}
  \corresp{\email{schu3@sheffield.ac.uk}},
  Ashley P. Willis\aff{1}
 \and Elena Marensi\aff{2}}

\affiliation{\aff{1} School of Mathematics and Statistics, University of Sheffield, Sheffield S3 7RH, UK
\aff{2} Department of Mechanical Engineering, University of Sheffield, Sheffield S1 3JD, UK}

\begin{document}
\maketitle

\begin{abstract}

It is well known that buoyancy suppresses, and can even laminarise, turbulence in upward heated pipe flow.  Heat transfer seriously deteriorates in this case.  A new DNS model is established to simulate 
flow-dependent heat transfer in an upward heated pipe.  The model shows good agreement with experimental results.  Three flow states are simulated for different values of the buoyancy parameter $C$: shear turbulence, laminarisation and convective turbulence.  The latter two regimes correspond to the heat transfer deterioration regime and the heat transfer recovery regime, respectively \citep{jackson2002influences,bae2005direct,zhang2020review}.
We confirm that convective turbulence is driven by a linear instability \citep{su2000linear} and that the deteriorated heat transfer within convective turbulence is related to a lack of rolls near the wall, which leads to weak mixing between the flow near the wall and the centre of the pipe.  Having surveyed the fundamental properties of the system, we perform a nonlinear nonmodal stability analysis, which seeks the minimal perturbation that triggers a transition from the laminar state.  Given the differences between shear and convective turbulence, we aim to determine how the nonlinear optimal (NLOP) changes as the buoyancy parameter $C$ increases.  We find that, at first, the NLOP becomes thinner and closer to the wall.  Most importantly, the critical initial
energy $E_0$ required to trigger turbulence keeps increasing, implying that attempts to artificially trigger it may not be an efficient means to improve heat transfer at larger C.
At 
$C=6$, a new type of NLOP is discovered, capable of triggering
convective turbulence from lower energy, but over a longer time.  
It is active only in the centre of the pipe.   
We next compare the transition processes, from linear instability and by the nonlinear nonmodal excitation.  At $C=4$, linear instability leads to a state that approaches a travelling wave solution or periodic solutions, while the minimal seed triggers shear turbulence before decaying to convective turbulence.  Deeper into the parameter space for convective turbulence, at $C=6$, the new nonlinear optimal triggers convective turbulence directly.  Detailed analysis of the periodic solution at $C=4$ reveals three stages: growth of the unstable eigenfunction, the formation of streaks, and the decay of the streaks. The stages of the cycle correspond to changes in the linear instability of the turbulent mean velocity profile. Unlike the self-sustaining process for classical shear flows, where the streak is disrupted via instability, here, decay of the streak is more closely linked to suppression of the linear instability of the mean flow, 
and hence suppression of the rolls.
Flow visualizations at $C$ up to $10$ also show similar processes, suggesting that the convective turbulence in the heat transfer recovery regime is sustained by these three typical processes. 
\end{abstract}


\section{Introduction}
\label{sec:headings}

 In isothermal pipe flow, the flow is driven by an external pressure gradient.  This is referred to as `forced' flow.  In a vertical configuration, however, buoyancy resulting from the expansivity of the fluid close to a heated wall can provide a force that partially or fully drives the flow, referred to as `mixed' or `natural convection' respectively.  The use of mixed convection is fundamentally important and practical in engineering applications, e.g. geothermal energy capture, nuclear reactor cooling systems, fossil fuel power plants, etc., and has been widely researched.  Heat transfer in the presence of buoyancy exhibits totally different characteristics for downward flow and upward flow. In a downward flow, the buoyancy acts as a drag force but always enhances the heat transfer. By contrast, in an upward flow, buoyancy assists the flow, but heat transfer can significantly deteriorate \citep{mceligot1970relaminarization,ackerman1970pseudoboiling,bae2005direct,wibisono2015numerical,zhang2020review}.  
When the heating parameter is gradually increased, 
 heat transfer first deteriorates, then recovers, and finally can approach as large values as for downward flow \citep{zhang2020review}. Interestingly, shear turbulence is gradually suppressed and even laminarised at lower Reynolds numbers \citep{bae2005direct,he2016laminarisation,chu2016direct,he2021turbulence}, then the flow enters a convective turbulence state when the buoyancy exceeds a critical intensity \citep{su2000linear,marensi2021suppression}.

Research on the phenomenon of laminarisation in mixed convection can be traced at least as far back as \cite{hall1969laminarization}, who provided a theoretical explanation of this phenomenon, suggesting that reduced shear stress in the buffer layer leads to a reduction or even elimination of turbulence production. This interpretation received wide acceptance. 
More recently, \cite{he2016laminarisation} modelled the buoyancy with a radially dependent body force added to isothermal flow, successfully reproducing the laminarisation phenomenon. They found that the body force causes little change to the key characteristics of turbulence, and proposed that laminarisation is caused by the reduction of the `apparent Reynolds number', which is calculated based only on the pressure force of the flow (i.e. excluding the contribution of the body force).  
In flows at supercritical pressure, 
the laminarisation and deterioration of heat transfer are reproduced successfully by \cite{bae2005direct,bae2006effects}. \cite{he2021turbulence} researched the laminarisation phenomenon in flows at supercritical pressure and established a unified explanation for the laminarisation mechanisms. It is thought to be due to the variations of thermophysical properties, buoyancy, and inertia, the latter of which plays a significant role in a developing flow.

Meanwhile, in isothermal flow, the laminarisation phenomenon
has been observed when the base velocity profile is flattened \citep{hof2010eliminating,kuhnen2018destabilizing,marensi2019stabilisation}. (The effect of buoyancy on the base velocity profile can be similar.) \cite{kuhnen2018destabilizing} proposed that the laminarisation is caused by reduced transient growth of small perturbations when the velocity profile is flattened. 
\cite{marensi2019stabilisation} considered the nonlinear stability of finite perturbations for
a series of flattened base velocity profiles in (isothermal) pipe flow, and found that flattening enhances the nonlinear stability of the laminar flow. 
Recently, machine learning has been used to explore laminarisation events in a reduced model of
wall-bounded shear flows, and it has been suggested that the collapse of turbulence is connected with the suppression of streak instabilities \citep{lellep2022interpreted}.  

There have been many studies of mixed convection, both experimental \citep{celataa1998upflow,wang2011investigation,jackson2013fluid,zhang2020review} and numerical 
\citep{you2003direct,poskas2012numerical,yoo2013turbulent,zhao2018direct},
but most of these works focus on statistical properties, e.g.\ the Nusselt number, Reynolds stress, mean velocity profile, the mean temperature profile and so on.  Few works pay attention to the dynamical characteristics of the flow, such as the transition between flow types
and the maintenance of convective turbulence.  In particular, the transition mechanisms for mixed convection appear to be quite different from those of isothermal flow.  Experimental results \citep{hanratty1958effect,scheele1960effect,ackerman1970pseudoboiling} have shown that a heated vertical pipe flow can go through a flow transition at Reynolds number as low as $30$. \cite{scheele1962effect} found  that the vertical heated upward flow in a pipe first becomes unstable when the velocity profiles develop an inflectional point. The flow develops regular and periodic motion after transition at rather a  low Reynolds number. Similar patterns have been also observed by \cite{kemeny1962combined}.  \cite{yao1987fully} confirmed the experimental observations that the flow in a heated vertical pipe is supercritically unstable. They found the bifurcated new equilibrium laminar flow is likely to be a double spiral flow. A weakly nonlinear instability analysis by \cite{rogers1993finite} revealed that the heated upward flow is supercritically unstable while the heated downward flow is potentially subcritically unstable.  \cite{su2000linear} systematically researched the linear stability of mixed-convection for upward and downward flow in a vertical pipe with constant heat flux, to explain multiple flow states that appeared in the experimental and numerical results. The calculation found that the first azimuthal mode is always the most unstable.  The Rayleigh-Taylor instability is operative when the Reynolds number is extremely low, while the opposed-thermal-buoyant instability is dominant when the Reynolds number is higher. The transition of a low Reynolds number mixed convection flow in a vertical channel has been investigated by \cite{chen2002direct} for K-type \citep{klebanoff1962three} and H-type \citep{herbert1983secondary} disturbances. 
It was found that the flow field bifurcates to a new quasi-steady nonlinear state after the initial transient period, for both types of perturbation. \cite{khandelwal2015weakly} developed a weakly nonlinear stability theory in terms of a Landau equation to analyse the nonlinear saturation of stably stratified nonisothermal Poiseuille flow in a vertical channel with respect to different fluids, i.e. mercury, gases, liquids, and heavy oils. A substantial enhancement in heat transfer rate was found for liquids and heavy oils from the basic state beyond the critical Rayleigh number. 
Recently, \cite{marensi2021suppression} also performed linear stability analysis for a vertical heated pipe flow, using the parameter $C$ to measure 
the buoyancy force relative to the force that drives the laminar isothermal shear flow. Their results show that the flow is not always linearly unstable at a strong buoyancy condition, and revealed that the heat transfer deterioration is caused by the suppression of rolls in the flow.  
This finding is in line with 
study of \cite{lellep2022interpreted}, as the suppression of rolls would occur when streak instabilities are suppressed.

Some works have considered the linear and weakly nonlinear stability of a vertical heated flow, and have shown the instability of convective turbulence.  However, the transition involving shear-driven turbulence
is fundamentally nonlinear, and multiple flow states can arise
at the same parameter values.
Further understanding is of great significance for heat transfer prediction and optimisation in engineering applications, and there remain open questions to be addressed, e.g.  How does the shear turbulence gradually disappear?  Is laminarisation similar to that in isothermal flow?  How is the convective turbulence triggered and sustained?  

To tackle these questions, we employ nonlinear nonmodal stability analysis \citep{pringle2010using,cherubini2011minimal,pringle2012minimal,cherubini2013nonlinear,marensi2019stabilisation} to seek the most efficient perturbation to trigger transition from the laminar state
in a vertical heated pipe.
The optimised perturbation is called the nonlinear optimal (NLOP), and the critical (i.e. lowest energy) NLOP that triggers turbulence is called
the `minimal seed' in isothermal flow. The magnitude and structure of the minimal seed reflects the nonlinear stability properties of the flow  \citep{marensi2019stabilisation}.  The Lagrange multiplier method has been widely used in fluid mechanics,
including the heat transfer optimisation approach proposed by \cite{guo2007entransy,motoki2018optimal}. 
We do not optimise heat transfer directly here, however, but 
seek the optimal flow perturbation that leads to transition
away from the low heat-transfer state of laminar flow.

 The plan of the paper is as follows. In \S 2, we present the new model for direct numerical simulation (DNS) of vertical heated pipe flow, and methods for the linear stability and nonlinear nonmodal stability analysis. In \S 3, we first show the results of DNS, then present the results of linear stability and nonlinear nonmodal stability analysis.  Next, we reveal the separate transition processes starting from the NLOP and unstable eigenfunctions, as well as the self-sustaining process in convective turbulence. Finally, the paper concludes with a summary in \S 4.

\section{Formulation}
\label{sec:headings}
    \subsection{Simulation of heated pipe flow}
    \thirdref{The upward flow through a vertical heated pipe flow is considered in this work.} Like several models, we assume there is a background temperature gradient along the axis of the pipe, e.g.\ \cite{yao1987fully,chen1996linear,chen2002direct,khandelwal2015weakly}. Different from many models, however, we suppose that the temperature gradient may vary in time due to changes in the way the fluid flows.
    We fix the difference between the bulk temperature of the 
    fluid and that of the wall, aiming to see directly how the buoyancy parameter affects the flow pattern, and hence the heat flux at the wall and the temperature gradient. 
    Our code is a modification to that of \cite{marensi2021suppression}, where
    the function of the temperature gradient was replaced by a spatially uniform heat sink.  That assumption had the advantage of providing a  simple analytic expression for the laminar state, here it is computed numerically.
    
    Let $\vec{x}=(r,\phi,z)$ denote cylindrical coordinates.
    We decompose the total temperature as  
\begin{equation}
\label{eq:tempexpansion}
   T_{tot}(\vec{x},t)=T_w(z,t) + T(\vec{x},t) - 
   2 T_b
\end{equation}
with wall temperature $T_w=a_{tot}(t)z+b$, 
\secondref{
where $a_{tot}(t)$ is the 
time-dependent axial temperature gradient, and $b$ is a constant reference temperature.   
Axial periodicity over a distance $L$ is assumed
for the temperature fluctuation field $T(\vec{x},t)$ and velocity field $\vec{u}_{tot}$.
Let $T_b=\langle T\rangle$, where the angle brackets denote the spatial average.
The factor $-2T_b$ has been inserted in  (\ref{eq:tempexpansion}) so that the  temperature fluctuations due to the heating from the wall are positive and largest at the wall: 
evaluating (\ref{eq:tempexpansion}) at the pipe radius $r=R$ gives $T|_{r=R}=2T_b$.
}

We take \secondref{$2T_b$}  
as our temperature scale for 
non-dimensionalisation, and assume that local fluctuations in  $T$ are more rapid than changes in the gradient $a_{tot}(t)$.  For the length and velocity scales, we non-dimensionalise with the radius $R$ and twice the bulk flow speed $2U_b$, the latter of which coincides with the isothermal laminar centreline speed. Throughout the rest of the text, dimensionless variables and equations are presented, except in the definition of dimensionless parameters.

Temperature fluctuations $T$ satisfy
        \begin{equation}
        	\frac{\partial T }{\partial t}+({\vec{u}_{tot}}\bcdot \vec{\nabla}) T =\frac{1}{RePr}\vec{\nabla} ^{2}T -{\vec{u}_{tot}}\bcdot{\hat{\vec{z}}}\,a_{tot}(t)
        	\label{T-total}
        \end{equation} 
with boundary condition $T=1$ at the wall.  A fixed bulk temperature,
$\langle T\rangle=1/2$, is maintained through adjustments in the gradient $a_{tot}(t)$, \firstref{which is
determined at each instant via the spatial average of the equation (see (\ref{at-equation}) below). }
The dimensionless parameters are the Reynolds and Prandtl numbers $Re=2U_bR/\nu$ and $Pr=\nu/\kappa$, where $\nu$ and  $\kappa$ are the kinematic viscosity and thermal diffusivity respectively. Under the Boussinesq approximation \citep{turner1979buoyancy}, the Navier--Stokes (NS) equations are
        \begin{equation}
        	\label{NSE-total}
        	\frac{\partial\vec{u}_{tot}}{\partial t}+(\vec{u}_{tot}\bcdot\vec{\nabla}) {\vec{u}_{tot}}=-\vec{\nabla} p_{tot} + \frac{1}{Re}\vec{\nabla}^{2} {\vec{u}_{tot}}+\frac{4}{Re}(1+\beta'(t)+C T )\hat{\vec{z}} \, ,
        \end{equation}
     with continuity equation
        \begin{equation}
       	\vec{\nabla} \bcdot \vec{u}_{tot}=0 \, ,
       \end{equation}
    and no-slip condition 
    \fourthref{$\vec{u_{tot}}=\vec{0}$}  
    at the wall. Here, $\beta'(t)$ is the excess pressure fraction, relative to isothermal laminar flow, required to maintain the fixed dimensionless mass flux, $\langle 
    {\vec{u}_{tot}}\cdot{\hat{\vec{z}}}
    \rangle=1/2$.  The dimensionless parameter
    \begin{equation}
    C=\frac{\mathrm{\it Gr}}{16\,\Rey} \, 
    \end{equation}
    measures the buoyancy force relative to the force that drives laminar isothermal shear flow, \secondref{ where 
    $\mathrm{\it Gr}={\gamma g(T|_{r=R}-T_b)(2R)^{3}}/{\nu ^{2}}$ 
    is the Grashof number, $\gamma$ is the coefficient of volume expansion, and $g$ is gravitational acceleration}.  For further details, see \cite{marensi2021suppression}.
   
    As the focus of this study is on the dynamics of perturbations from the laminar solution, we decompose the variables,
    ${\vec{u}_{tot}}=\vec{u}_0+\vec{u}$, 
    $p_{tot}=p_0+p$, 
    $1+\beta'(t)=1+\beta_{0}+\beta(t)$, 
    $T=\Theta_0+\Theta$, 
    $a_{tot}(t)=a_{0}+a(t)$,
    where subscript $0$ tags the laminar solution, \secondref{$\vec{u}=(u_r,u_{\phi},u_z)$}.
    The laminar velocity $\vec{u}_0=u_0(r)\,\hat{\vec{z}}$
    satisfies
            \begin{equation}
        	\label{NSE-laminar}
        	-\vec{\nabla} p_{0} + \frac{1}{Re}\vec{\nabla}^{2} {\vec{u}_{0}}+\frac{4}{Re}(1+\beta_0+C \Theta_0)\hat{\vec{z}}=0 \, .
        \end{equation}
    Subtracting (\ref{NSE-laminar}) from (\ref{NSE-total}), we obtain  
        \begin{equation}
        	\label{NSE-perturbation}
        	\frac{\partial \vec{u}}{\partial t}+u_0\frac{\partial \vec{u}}{\partial z}+u_{r}\,\frac{du_0}{dr}\, \hat{\vec{z}}+(\vec{u}\bcdot \vec{\nabla}) \vec{u} =-\vec{\nabla}  p + \frac{1}{Re}\vec{\nabla} ^{2}\vec{u}+\frac{4}{Re}(C \Theta+\beta(t))\hat{\vec{z}}.
        \end{equation}
   \firstref{Taking the spatial average of the $z$-components of 
   (\ref{NSE-laminar}) and
   (\ref{NSE-perturbation}) gives equations for $\beta_0$ and $\beta(t)$
   that fix $\langle u_z \rangle = 0$:
   \begin{equation}
         \label{betat-equation}
            \beta_0 = 
            -\frac{1}{2} 
            \left( \left.\frac{\partial u_0}{\partial r}\right|_{r=1}+C\right), \qquad
        	\beta(t) =   
         -\frac{1}{2} \left.\frac{\partial (u_z)_{00}}{\partial r}\right|_{r=1},
         \end{equation}
   where $(\cdot)_{00}$ 
   corresponds to averaging over $\phi$ and $z$. }
   The laminar temperature profile $\Theta_0(r)$ satisfies
        \begin{equation}
        	\label{T-laminar}
        	\frac{1}{RePr}\vec{\nabla} ^{2}\Theta_0 =\vec{u}_{0}\bcdot {\hat{\vec{z}}}a_{0} \, .
        \end{equation}
    Subtracting (\ref{T-laminar}) from (\ref{T-total}), we obtain
        \begin{equation}
        	\label{T-perturbation}
        	\frac{\partial \Theta }{\partial t}+u_0\frac{\partial \Theta }{\partial z}+u_r\frac{d\Theta_{0}} {dr} + (\vec{u}\bcdot\vec{\nabla}) \Theta =\frac{1}{RePr}\vec{\nabla} ^{2}\Theta -u_z a_{0}-(u_0+u_z)a(t).
        \end{equation}
        \firstref{
        Taking the spatial averages of 
        (\ref{T-laminar}) and 
        (\ref{T-perturbation}) gives equations that determine $a_0$ and the value of $a(t)$ required to fix
        $\langle\Theta\rangle=0$:
         \begin{equation}
         \label{at-equation}
            a_0 = 
            \frac{4}{RePr}
            \left.\partial_{r}\Theta_{0}\right|_{r=1},
             \qquad          
        	a(t)= 
            \frac{4}{RePr}
            \left.\partial_{r}(\Theta)_{00}\right|_{r=1} \, .
         \end{equation}
         }
         
\subsection{Linear stability analysis}
    Arnoldi iteration is employed to calculate the leading eigenvalues and eigenfunctions of the laminar solution using our simulation code. 
    Considering a small perturbation $\vec{u}$ to the laminar solution $\vec{u}_0$, \fourthref{so that the nonlinear terms of (\ref{NSE-perturbation}) may be ignored,} the linearised system may be written
    in the form
         \begin{equation}
             \label{eq:lin}
     	     \partial _{t}\vec{u}=A(u_0)\,\vec{u}\,.
         \end{equation}
    Integrating over a period $\Topt$, eigenfunctions of (\ref{eq:lin})
    with growth rate $\sigma$ satisfy the exponentiated eigenvalue
    problem
    \begin{equation}
        \vec{u}(\Topt)
        = \vec{u}(0) + \int_0^{\Topt} A(u_0)\, \vec{u}(t) \, \mathrm{d}t
        \,=\, B(u_0)\,\vec{u}(0) 
        \,= \, \mathrm{e}^{\sigma \Topt}\vec{u}(0) 
        \, .
    \end{equation}
    The Arnoldi method only requires the result of the calculations of multiplies by $B$ with given $\vec{u}$, i.e.\ the result of time integration of a $\vec{u}$ over the period $\Topt$.
    Given a starting $\vec{u}$, the method seeks eigenvectors in the Krylov subspace $\mathcal{K}=\mathrm{span} \{\vec{u}, B\vec{u}, B^2\vec{u},...\}$, using Gram-Schmidt orthogonalisation to improve the numerical suitability of this basis set. 
   Using $f(\vec{u})$ to denote the result of time integration of a perturbation $\vec{u}$ from the laminar state,
         we may make the approximation
         \begin{equation}
         	 B(u_0)\, \vec{u}\approx \frac{1}{\epsilon }
            \left\{ 
               f(\vec{0}+\epsilon \vec{u})-f(\vec{0})
            \right\}
            ~=~ \frac{1}{\epsilon}f(\epsilon\vec{u})\, ,
         \end{equation}
for some small value $\epsilon$. \firstref{
We take $\epsilon$ such that $\epsilon\|\vec{u}\|=10^{-6}\|\vec{u}_0\|$, i.e.\ the perturbation is 6 orders of magnitude smaller than the laminar state, and nonlinear terms are 12 orders smaller.  The choice of $\Topt$ does not affect evaluation of the real part of $\sigma$, which determines stability, but a large $\Topt$ can cause aliasing of its complex part.  The period $\Topt$ is therefore chosen to be small, but not too small, as the number of Arnoldi iterations required is usually inversely proportional to $\Topt$, and many iterations would require storage of a large $\mathcal{K}$.  With the aim of keeping the calculation to around 50 Arnoldi iterations,
typically $\Topt\approx 10$.
}

%

\subsection{Nonlinear nonmodal stability analysis}
    When the linear system is stable, we must appeal to nonlinear dynamics to interpret transition. The NLOP \citep{pringle2010using,pringle2012minimal}  
    is the optimal perturbation that achieves the largest energy growth, and when just large
    enough to trigger transition, is called 
    the minimal seed.
    
    The NLOP can be found by the Lagrange multiplier technique; for details see \cite{pringle2012minimal}.
    To reduce complexity, we consider only the optimal velocity perturbation, and set the initial temperature perturbation to zero.
    In this case, the Lagrangian is defined by
    \begin{equation} 
        \begin{split}
        		\mathcal L ~=~ \langle\frac{1}{2}(\vec{u}(x,\Topt))^{2}\rangle -\lambda_0 (\frac{1}{2}\langle(\vec{u}(x,0))^{2}\rangle-E_{0})-\int_{0}^{\Topt}\langle \vec{v} \bcdot \mathrm{NS}(\vec{u})\rangle \, \mathrm{d}t-\\ \int_{0}^{\Topt}\langle\Pi \, \vec{\nabla}\bcdot \vec{u}\rangle \, \mathrm{d}t -\int_{0}^{\Topt}\langle\pi \, \mathrm{Tem}(\Theta)\rangle \, \mathrm{d}t-\\ \int_{0}^{\Topt} 
                \Gamma \langle \vec{u}\bcdot \hat{\vec{z}} \rangle\, \mathrm{d}t-\int_{0}^{\Topt} Q \langle \Theta \rangle\, \mathrm{d}t \, ,
        \end{split} 
    \end{equation}
   \firstref{where $\lambda_0$, $\Pi(\vec{x},t)$, $\pi(\vec{x},t)$, $\Gamma(t)$, $Q(t)$ and \secondref{$\vec{v}(\vec{x},t)=(v_r,v_{\phi},v_z)$} are Lagrange multipliers. The first term, the perturbation energy at final time $t=\Topt$, is the objective function to be maximised.  The second term is the constraint of fixed amplitude for the initial perturbation. Then, the velocity perturbation $\vec{u}$ is constrained to satisfy the Navier--Stokes equation $\mathrm{NS}(\vec{u})$ and the continuity equation from $t=0$ to $t=\Topt$, while the temperature perturbation satisfies the temperature equation $\mathrm{Tem}(\Theta)$ from $t=0$ to $t=\Topt$. The velocity and temperature must also satisfy fixed mass flux and fixed heat flux, represented by the last two terms.}
   Taking variations of $\mathcal L$ with respect to each variable and setting them equal to zero, we obtain the following set of Euler--Lagrange equations.  The adjoint Navier--Stokes, temperature equation and continuity equations are
     \begin{equation} \label{eq:adjnu}
        	\begin{split} 
        		\frac{\delta \mathcal L} {\delta \vec{u}}=\frac{\partial \vec{v}}{\partial t}+u_0\frac{\partial \vec{v}}{\partial z}-v_{z}{u'_0}{\hat{\vec{r}}}+ \vec{\nabla}\times(\vec{v}\times \vec{u})- \vec{v}\times\vec{\nabla} \times \vec{u} +\vec{\nabla} \Pi + \\ \frac{1}{Re}\vec{\nabla} ^{2}\vec{v}-\pi\Theta'_{0}\hat{\vec{r}}-\pi\vec{\nabla}\Theta-\pi (a(t)+a_0)\hat{\vec{z}}-\Gamma \hat{\vec{z}}=0 \, ,
        	\end{split} 
     \end{equation}
        \begin{equation} \label{eq:adjT}
        	\frac{\delta \mathcal L}{\delta \Theta}=\frac{\partial \pi}{\partial t}+u_0\frac{\partial \pi}{\partial z}+\frac{4}{Re} v_z C+\vec{u}\bcdot \vec{\nabla}\pi + \frac{1}{Re Pr}\vec{\nabla}^2\pi-Q=0 \, ,
        \end{equation}
        \begin{equation}
         \label{continuty-v}
        	\frac{\delta \mathcal L}{\delta p}=\vec{\nabla} \bcdot \vec{v} =0 \, ,
        \end{equation} 
    where primes indicate the radial derivative.
    The compatibility conditions
    are given by
        \begin{equation} \label{eq:compatu}
        	\frac{\delta \mathcal L}{\delta \vec{u}(\vec{x},\Topt)}=\vec{u}(\vec{x},\Topt)-\vec{v}(\vec{x},\Topt)=0
        \end{equation} 
        \begin{equation} \label{eq:compatT}
        	\frac{\delta \mathcal L}{\delta \Theta(\vec{x},\Topt)}=-\pi(\vec{x},\Topt)=0.
        \end{equation} 
 and the optimality condition is
        \begin{equation} \label{eq:optimality}
        	\frac{\delta \mathcal L}{\delta {\vec{u}(\vec{x},0)}}=-\lambda_0 \vec{u}(\vec{x},0)+\vec{v}(\vec{x},0)=0.
        \end{equation}
    \firstref{
    Optimisation is performed iteratively from a starting velocity field $\vec{u}(\vec{x},0)^{(0)}$, where the superscript indicates the iteration.
    Given a field $\vec{u}(\vec{x},0)^{(j)}$, the Navier--Stokes equations are integrated to time $\Topt$.  The compatibility conditions (\ref{eq:compatu}) and (\ref{eq:compatT}) provide end conditions for the backwards integration of (\ref{eq:adjnu}) and (\ref{eq:adjT}) to time $0$.  Like $p$ and $\beta$, the Lagrange multipliers $\Pi$ and $\Gamma$ are used to ensure that $\vec{\nu}(\vec{x},t)$ is also divergence-free and has zero flux.  $Q$ is used to fix that $\langle \pi\rangle=0$.  Finally, once $\vec{\nu}(\vec{x},0)$ has been calculated, the optimality condition 
    (\ref{eq:optimality}) may be evaluated.       }
    As the optimality condition is not satisfied automatically, $\vec{u}(\vec{x},0)$ is moved in the direction of $ \delta\mathcal L/\delta \vec{u}(\vec{x},0)$ towards achieving a maximum for $\mathcal{L}$, where $ \delta\mathcal L/\delta \vec{u}(\vec{x},0) $ should vanish. 
    The update to $\vec{u}(\vec{x},0)^{(j)}$ each iteration is 
    \begin{equation}
        	\vec{u}(\vec{x},0)^{(j+1)}=\vec{u}(\vec{x},0)^{(j)}+\fourthref{\epsilon_0}\frac{\delta \mathcal L}{\delta \vec{u}(\vec{x},0)^{(j)}} \, ,
        \end{equation}
    where \fourthref{$\epsilon_0$} is a small value, controlled using a procedure described in \cite{pringle2012minimal}, 
    and
    $\lambda_0$ is adjusted to set $\langle[\vec{u}(\vec{x},0)^{(j+1)}]^2\rangle=2\,E_0$.  

\subsection{Time integration code}
    \label{sec:timeint}
    The calculations are carried out by the open-source code openpipeflow.org \citep{willis2017openpipeflow}. 
    Variables
    are discretised on the domain $\left\{ r,\phi,z \right\}=[0,1]\times[0,2\pi]\times[0,2\pi/k_0]$, where $k_0=2\pi/L$, using Fourier decomposition in the azimuthal and streamwise directions and finite difference in the radial direction, 
    \begin{equation} \label{eq:discr}
      \left\{\vec{u},p,\Theta\right\}(r_s,\phi,z)=\sum_{k<|K|} \sum_{m<|M|}\left\{\vec{u},p,\Theta \right\}_{skm}e^{i(k_0kz+m\phi)} \, ,
        \end{equation}
    where $s=1,...,S$ and the radial points are clustered near the wall. Temporal discretisation is via a second-order predictor-corrector scheme, with an Euler predictor and a Crank-Nicolson corrector applied to the nonlinear terms.  
    The laminar solution is quickly calculated by eliminating 
    azimuthal and axial variations using a 
    resolution $S=64, M=1, K=1$. 
    For nonlinear calculations, as the adjoint optimisation requires forward and backward time integrations for each iteration, it is computationally expensive. 
    To keep calculations manageable, a Reynolds number $Re=3000$ is adopted for the nonlinear nonmodal stability analysis with a domain of length $L=5D$.
    We use $S=64, M=48, K=42$
    and a time step $\Delta t=0.01$. 
    
\section{Results}
\label{sec:headings}
\subsection{Flow regimes}

\begin{figure}
        \centering
        \includegraphics[angle=0,width=0.9\textwidth]{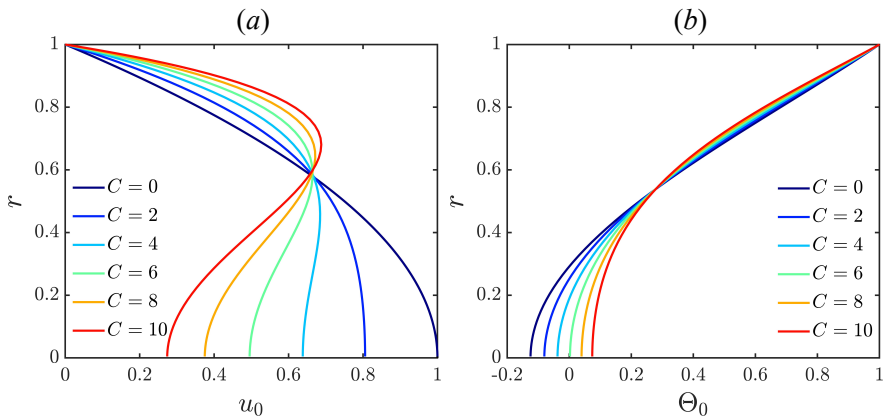}
        \caption{($a$) the laminar velocity profile, ($b$) the laminar temperature profile.  Laminar profiles are independent of Reynolds number.}
        \label{laminar-solution}
     \end{figure}

\begin{figure}
       \includegraphics[angle=0,width=1\textwidth]{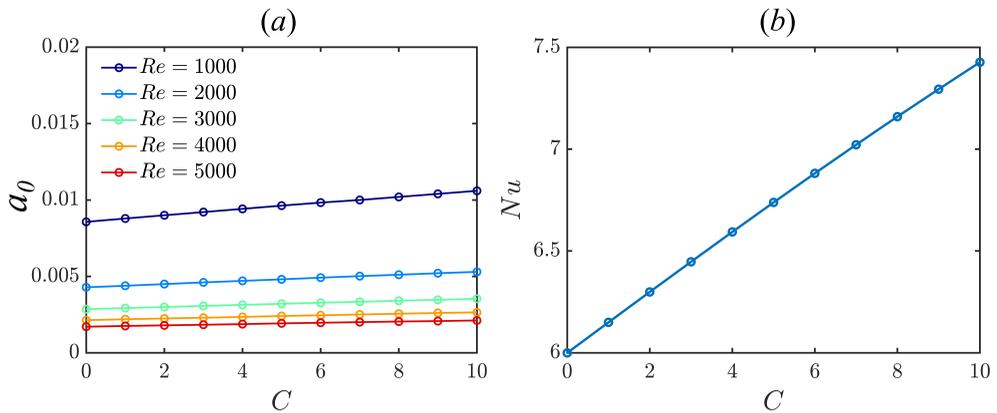}  
       \caption{($a$) Temperature gradient of laminar flow $a_0$ as a function of $C$ for several $\Rey$. ($b$) Nusselt number as a function of $C$. }
      \label{laminar-tem-gradient}
    \end{figure}

    Figure \ref{laminar-solution} shows how the laminar velocity and temperature profile changes with $C$. It is observed that the velocity profile becomes flattened, then 
    is well known to become `M-shaped' at larger values of $C$. 
    The temperature profile also changes, but not so strongly.  The boundary temperature gradient of the laminar state is shown in figure \ref{laminar-tem-gradient}$(a)$.  
    The laminar temperature gradient increases 
    with slightly with $C$ and decreases substantially
    with $\Rey$,
     as the higher flow rate carries heat out of the system more rapidly. 
    The Nusselt number is defined 
        \begin{equation}
         \mathrm{\it Nu}
         =\frac{2R\, q_w}{\lambda \, (\firstref{T|_{r=R}-T_b})}\,,
        \end{equation}
    where $\lambda$  is the thermal conductivity of the fluid, and $q_w$ is net wall heat flux.  The laminar Nusselt number increases with a larger value of $C$, see figure \ref{laminar-tem-gradient}$(b)$, but it is independent of the Reynolds number \citep{su2000linear}. 
    
\begin{figure}
          	\centering
          	\includegraphics[angle=0,width=1\textwidth]{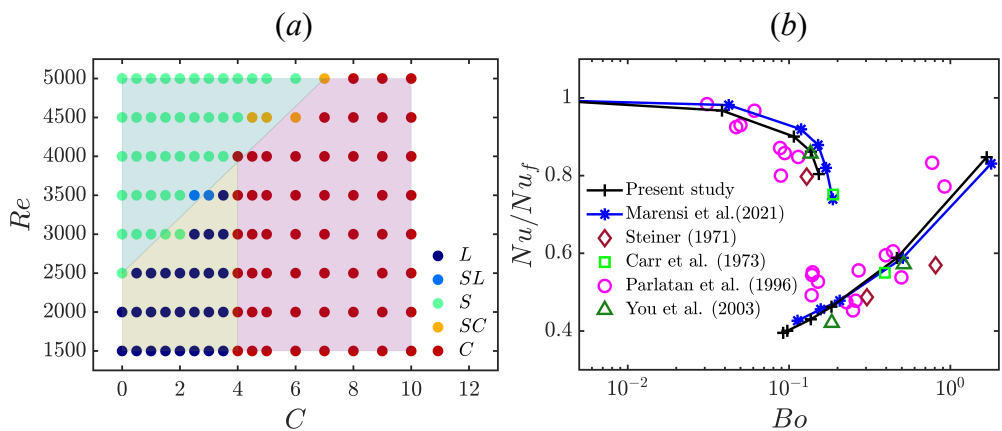}  
          	\caption{ $(a)$ Regions of laminar flow (L), shear  turbulence (S) and convective turbulence (C). SL and SC indicate
           that the flow may be in either of the two states. 
           $(b)$ Change in heat flux, normalised by that for the isothermal state ($C\to 0$), as a function of $Bo = 8\times 10^4\,(8\,Nu\,Gr)/(\Rey^{3.425}Pr^{0.8})$. 
           Present data from simulations at $Re = 5300$, $Pr = 0.7$ and various $Gr = 16\,\Rey\,C$. The upper and lower branches correspond to shear turbulence and convective turbulence respectively.}
          	\label{SCL-boundary}
     \end{figure}

    As discussed in the introduction, there are three typical flow states: laminar flow, shear turbulence, and convective turbulence (convection-driven turbulence). 
    We have performed a suite of simulations at different Reynolds number $\Rey$ and buoyancy parameter $C$ and the observed flow regime 
    is shown in
    figure \ref{SCL-boundary}$(a)$.  The diagram is consistent with figure 7 of \cite{marensi2021suppression}. 
    Close to the boundary, 
    where the laminar or convective turbulence
    regimes meet shear turbulence, the flow
    can be found in either state.  Nevertheless, the
three typical flow regimes can be identified clearly.  In the iso-thermal case ($C=0$), shear-turbulence appears for $Re\gtrsim 2000$ \citep{avila2011onset,avila2023transition}, but increasing the effect of buoyancy, at lower Reynolds numbers, the flow first laminarises then transitions to convective turbulence. At a higher Reynolds number, the laminarisation regime becomes narrower and finally disappears. Convective turbulence first appears for $C\gtrsim 4$ at lower Reynolds number, as reported by \cite{scheele1960effect,scheele1962effect,yao1987fully}. But at high Reynolds number the shear-driven state persists to larger $C$ and 
    crosses the critical $C$ observed at low Reynolds number, setting up a direct transition from the shear-driven turbulence to convective turbulence as $C$ is increased.
    \firstref{The flow states at $\Rey$ 
    larger than 5300 have not been computed in the 
    present work, but this direct transition is known
    to occur at much larger $\Rey$, and a collapse in experimental data is observed when presented in terms of the buoyancy number $Bo= 8\times 10^4\, (8\,Nu\,Gr)/(\Rey^{3.425}Pr^{0.8})$, where $Gr = 16\,\Rey\,C$, as seen in figure \ref{SCL-boundary}(b). }
     
    Numerical results are quantitatively validated by comparison with experiments in figure \ref{SCL-boundary}$(b)$, which also includes the numerical results of \cite{marensi2021suppression}. 
    The strong heat flux for smaller values of buoyancy number $Bo$ is brought about by shear turbulence, but deteriorates as $Bo$ is increased when buoyancy spresses the shear turbulence.  When $Bo$ exceeds a certain value, the flow switches from shear turbulence to convective turbulence, accompanied by a sudden  drop in heat transfer. However, the heat transfer eventually recovers with further increase of $Bo$. The captured heat transfer features are consistent with the results reported by  \cite{bae2005direct,yoo2013turbulent}.  
    In the present model, the velocity field is allowed to affect the heat flux at the wall; the model and code are a minor update 
    to that of \cite{marensi2021suppression}, which assumed a constant boundary temperature with a spatially uniform heat sink term.
    The uniform sink has the advantage that there is a simple analytic expression for the laminar base flow, but the axial temperature gradient which leads to the spatially dependent heat sink term is expected 
    to be closer to the real system.  Compared with \cite{marensi2021suppression}, it is found that the present model provides a small improvement in agreement with experimental data in the shear-turbulence regime, while in the convective turbulence regime, the two models give similar results.  

\begin{figure}
          	\centering
          	\includegraphics[angle=0,width=1\textwidth]{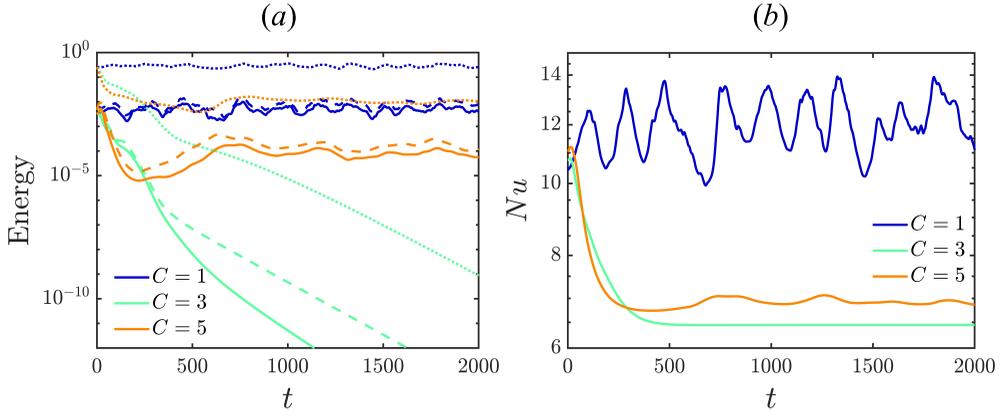}  
          	\caption{
           Simulation at $Re=3000$ starting from the same initial condition at $C=1, 3, 5$.
           ($a$) Velocity perturbation energy. 
           The \fourthref{solid, dash and dotted} line represent the energy of three components of velocity, marked as $E(u_r)$, $E(u_{\phi})$ and $E(u_z)$, respectively. 
                      ($b$) Nusselt number.}
          	\label{Flow-states}
    \end{figure}

    Starting from isothermal flow ($C=0$) at $Re=3000$,
    figure \ref{Flow-states} shows the effect of switching on the buoyancy for $C=1,3,5$. When the buoyancy is weak, at $C=1$, in figure \ref{Flow-states}($a$), we find that the amplitude of all three components maintain a high amplitude, and the flow is essentially unchanged from the isothermal case.
    For $C=3$, the velocity perturbation energy decays quickly, indicating the occurrence of laminarisation. 
    At $C=5$, initially a similar energy drop is also observed, but it stops at a much lower energy level and fluctuates with a low frequency, which is typical of the convective turbulent state. Heat transfer for the laminar and convective turbulent state is severely reduced, 
    seen in figure \ref{Flow-states}($b$), where the Nusselt number drops to approximately half of that for shear turbulence.  The convective turbulent state at these parameters has a Nusselt number that is not much greater than for the laminar state.
    
\begin{figure}
          	\centering
          	\includegraphics[angle=0,width=0.9\textwidth]{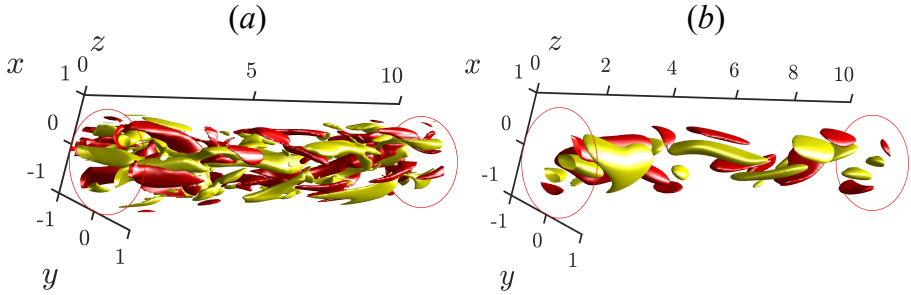}  
          	\caption{The isosurface of streamwise vorticity at ($a$) shear turbulence $(C=1)$ and ($b$) convective turbulence $(C=5)$ at $Re=3000$; red/yellow are $30\%$ of the min/max streamwise vorticity.}         
          	\label{isoUz-shear-convective}
    \end{figure}

\begin{figure}
          	\centering
          	\includegraphics[angle=0,width=1\textwidth]{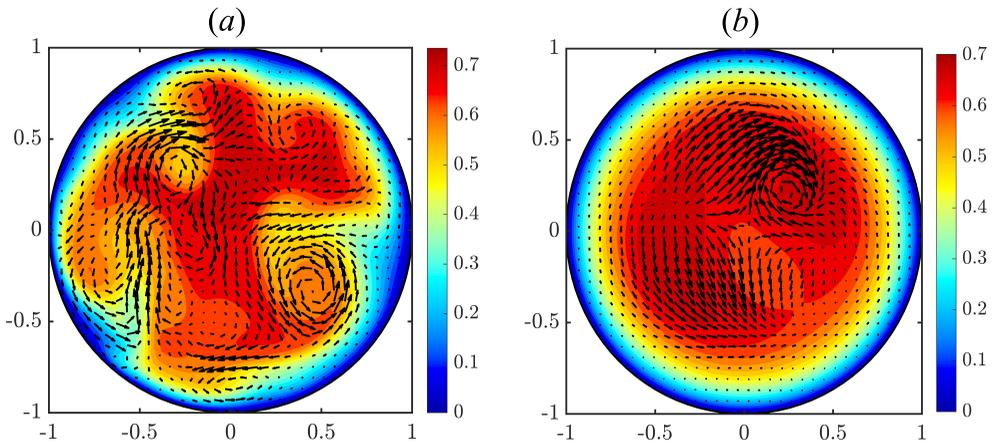}  
          	\caption{The cross-section of ($a$) shear turbulence $(C=1)$ and ($b$) convective turbulence $(C=5)$ at $Re=3000$, the contours are colored by streamwise total velocity, and the arrows represent cross-stream components. The largest arrow has magnitude $5.6e-3$ in $(a)$ and $1.93e-4$ in $(b)$.
           }    
                \label{contour-shear-convective}
    \end{figure}

    Isosurfaces of the streamwise vorticity are plotted in figure \ref{isoUz-shear-convective} for shear turbulence $(C=1)$ and convective turbulence $(C=5)$. Shear turbulence has complicated streamwise vortices that fill the whole pipe,  while the convective turbulence exhibits more organised vortex structures that are mainly concentrated around the middle of the pipe.  The location of the structures can be seen more clearly in the cross-sections of figure \ref{contour-shear-convective}. In shear turbulence, 
    there are abundant near-wall streamwise vortices, which lift up low-speed streaks, and enhance the mixing of fluid in the pipe. 
    In convective turbulence, 
    the perturbations are found around the centre of the pipe and the near-wall flow is almost steady. 
    There are almost no near-wall rolls and streaks observed. Further visualisations of flow at larger $C$ show similar characteristics.  
    The self-sustaining process of shear turbulence,
    reliant on near-wall vortices for the creation of streaks,
    has been destroyed in convective turbulence.  
    The lack of streamwise vortices significantly weakens the interaction between fluid near the wall and center, and heat transfer is therefore seriously damaged.
    
 \subsection{Linear stability results}
\begin{figure}
          	\centering
          	\includegraphics[angle=0,width=1\textwidth]{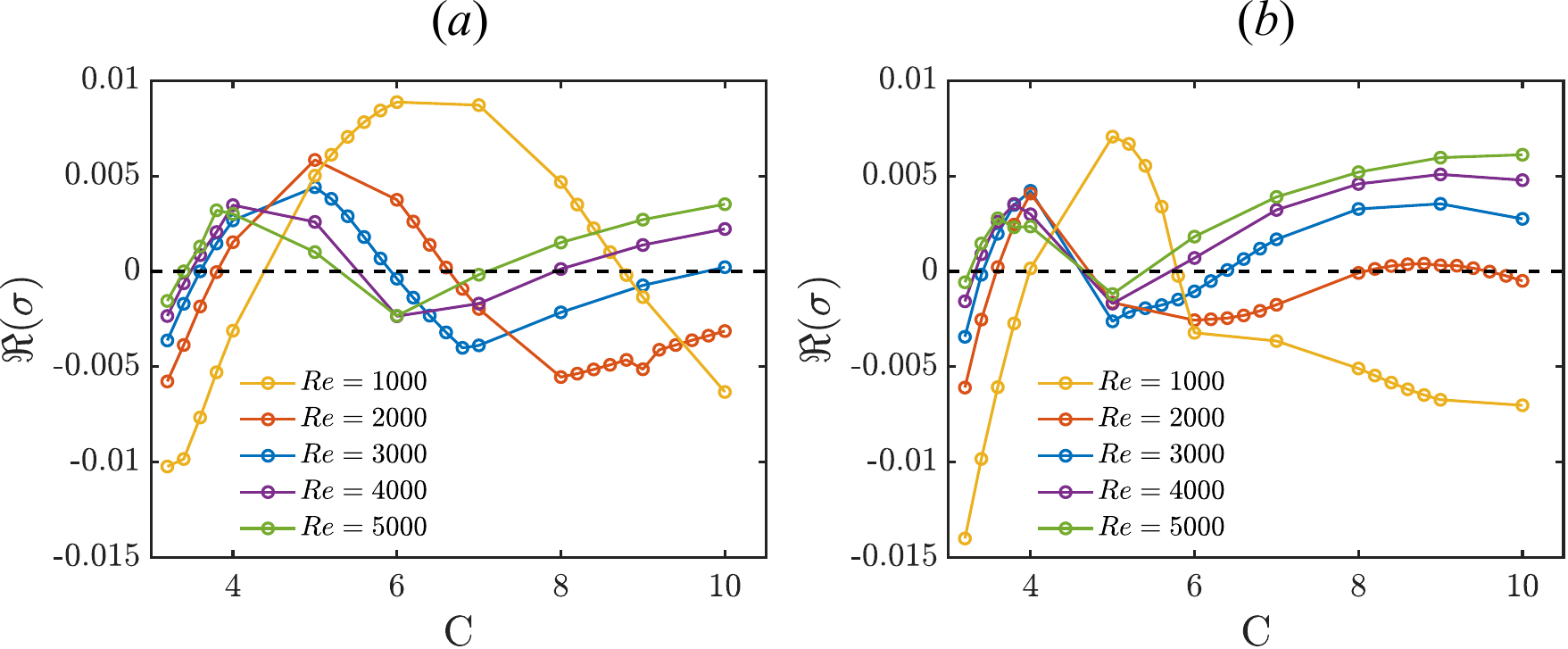}  
          	\caption{
           Growth rate of perturbations of different streamwise wave numbers ($a$) $m=1, k=1$ ; ($b$) $m=1, k=2$ versus buoyancy parameter $C$ at several Reynolds numbers.}        
          	\label{linear-stability}
        \end{figure}

\begin{figure}
          	\centering
          	\includegraphics[angle=0,width=1\textwidth]{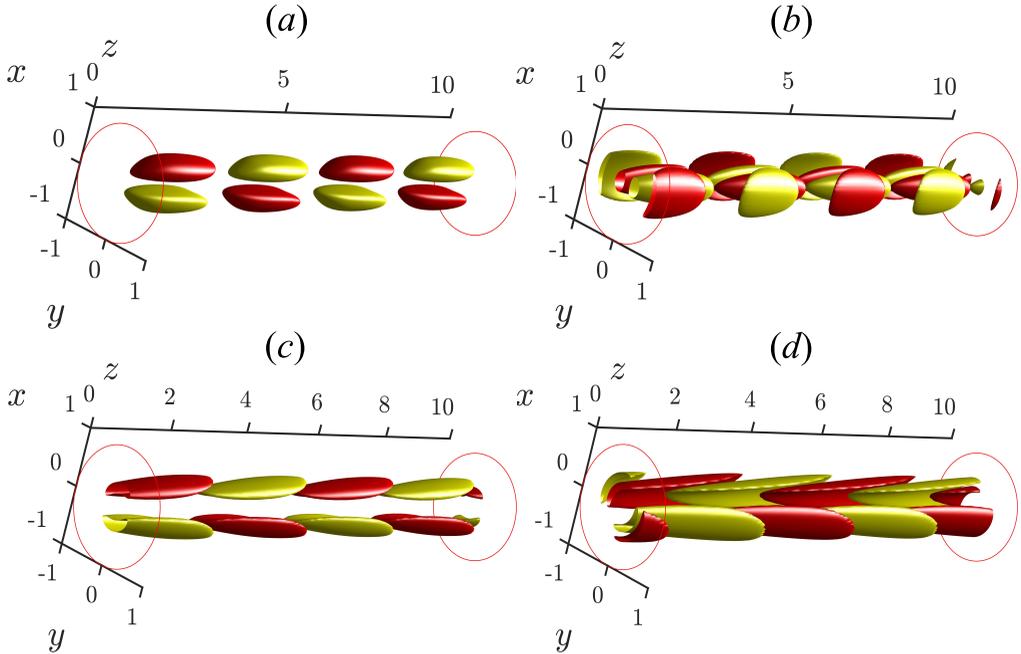}  
          	\caption{Isosurfaces of eigenfunctions for wave number $k=2, m=1$ at $Re=3000$.
 ($a,b$) First branch at $C=4$, $(c,d)$ second branch at $C=10$. 
 $(a,c)$ Streamwise velocity perturbations, $(b,d)$ streamwise vorticity. 
 (Red/yellow are 30$\%$ of the min/max values.)}      
          	\label{unstable-mode}
        \end{figure}

  This section examines the change of linear stability of the laminar solution at different $C$, revealing how the buoyancy affects the dynamics near the laminar solution. 
    Time stepping is combined with Arnoldi iteration to calculate the leading eigenvalues and  eigenfunctions for small perturbations about the laminar state.
A resolution of $S=64, M=10, K=10$ is used.
\firstref{A finer radial resolution does not produce a noticable difference in the results,} and the 
low $K$ and $M$ is sufficient to pick the most unstable axial and azimuthal modes for the given domain ($L=5D$). 
It is found that the most unstable mode is 
always of azimuthal wavenumber $m=1$ and is 
usually of wave number $k=1$ or $k=2$.
These observations are consistent with results of \cite{scheele1960effect,su2000linear,marensi2021suppression}. 
The real part of the eigenvalue of the unstable modes is shown in figure \ref{linear-stability}. The instability first appears near $C=4$,  consistent with the appearance of convective turbulence, verifying  that convective turbulence is caused by linear instability \citep{yao1987fully,rogers1993finite,marensi2021suppression}.
    Two branches of unstable mode lead to the instability of the flow,
    \secondref{
    apart from at the lowest $\Rey$.
    For both $k=1$ and $k=2$ cases, the first branch
    enters just before $C=4$ but stabilises by 
    $C\approx 6$.  A second branch then takes 
    over for $C\gtrsim 6$.  
    Comparing eigenvalues between the $k=1$ and $k=2$
    cases, the first branch with $k=1$ leads at lower $C$ and the second branch with $k=2$ leads at larger $C$.}
    Interestingly, the flow is not always linearly unstable for $C>4$, as there is a  window of stability for $C\approx 6$ for a range of $\Rey\approx 2000-4000$.  This was also reported for the model of \cite{marensi2021suppression}.  
    In particular, at $Re=3000$ the flow is linearly stable for $C=6$ --- we will take advantage of this for the nonlinear stability analysis of the following subsection \S\ref{subsect:nonlinstab}.
    Eigenfunctions of the two branches of unstable modes with wave number $k=2$, $m=1$  are shown in figure \ref{unstable-mode} with isosurfaces of streamwise velocity and streamwise vorticity.  The eigenfunctions are consistent with the results of \cite{khandelwal2015weakly,marensi2021suppression}.  The perturbation is mainly active in the centre of the pipe, in good agreement with the observations of \cite{su2000linear,marensi2021suppression}.

    Linear stability analysis results have confirmed that convective turbulence is triggered by linear instability, thus it mainly appears at $C>4$ as shown by the magenta shadowed region in figure \ref{SCL-boundary}$(a)$. However, at a high Reynolds number, the critical $C$ for transition is pushed back to larger $C$. This indicates competition between shear turbulence and convective turbulence. 
    For he limit of the shear-turbulence region shown in figure \ref{SCL-boundary}$(a)$,
    various possible explanations have been proposed.
    \cite{he2016laminarisation} linked the buoyancy-modified flow
    to a partner isothermal flow at a different (lower) `apparent Reynolds number', 
    based on an apparent friction velocity associated with only the pressure force of the flow (i.e.\ excluding the contribution of the body force). They found that the buoyancy can reduce the apparent Reynolds number, thereby suppressing turbulence. \cite{kuhnen2018destabilizing} suggested that it is the flattened base velocity profile, see figure \ref{laminar-solution}, which reduces the transient growth of streaky perturbations, and thereby causes the laminarisation. \cite{marensi2019stabilisation} did a nonlinear non-normal stability analysis of the flattened base velocity profile, suggesting the nonlinear stability of the flattened
    velocity profile is enhanced, i.e.\ a larger amplitude perturbation
    is required to trigger turbulence. 
    \cite{lellep2022interpreted} used a machine learning method and linked the collapse of turbulence to the suppression of streak instability.  We hold a similar view to \cite{lellep2022interpreted} --- through linear and nonlinear optimisation
     \citep{CSJ24}, we have found that a body force which flattens the base velocity profile always 
     produces streaks with a more stable shape.
     Further numerical experiments have been carried out, imposing a body force which flattens the base velocity profile at particular times during the relative periodic orbit, i.e.\ only during the time interval of the formation of streaks, or the interval for the regeneration of rolls.  Only when targeting the formation of streaks was the self-sustaining process 
 disrupted, supporting that the flattened profile suppresses turbulence by the stabilisation of streaks. With the increase of Reynolds number, the streaks formed by the flattened base velocity profile will recover the instability to sustain turbulence. Therefore, the laminarisation regime gradually disappears at a higher Reynolds number, see the green shadowed region in figure \ref{SCL-boundary}$(a)$. 
    
\subsection{Nonlinear nonmodal stability analysis} \label{subsect:nonlinstab}

     \firstref{Isothermal pipe flow is linearly stable, and nonlinear dynamics must considered when examining transition.  As $C$ is increased, 
     we have just seen that linear instability arises, but also that 
     there is a range of linear stability even in the convective regime.  } 
     In this section, we consider
     a perturbation of given finite amplitude $A_0$ that grows most over a time $\Topt$.  This perturbation is the nonlinear optimal 
     (NLOP).  When $A_0$ is just large enough to trigger turbulence,
     the NLOP is known as the `minimal seed'. 
     We seek to see how the NLOP is affected by buoyancy, and focus 
     on $Re=3000$ for $C$ in the range $0-6$.
     Nonlinear flows at $C=0,1,2$ are in the shear turbulence regime, while flow at $C=3$ returns to the laminar regime. Flows at $C=4,5,6$ are in the convective turbulence regime.  
     Note, however, that only the laminar flows at $C=4$ and $5$ are linearly unstable, while at $C=6$ it is linearly stable yet well within the convective regime.

\begin{figure}
          \centering
          \includegraphics[angle=0,width=1\textwidth]{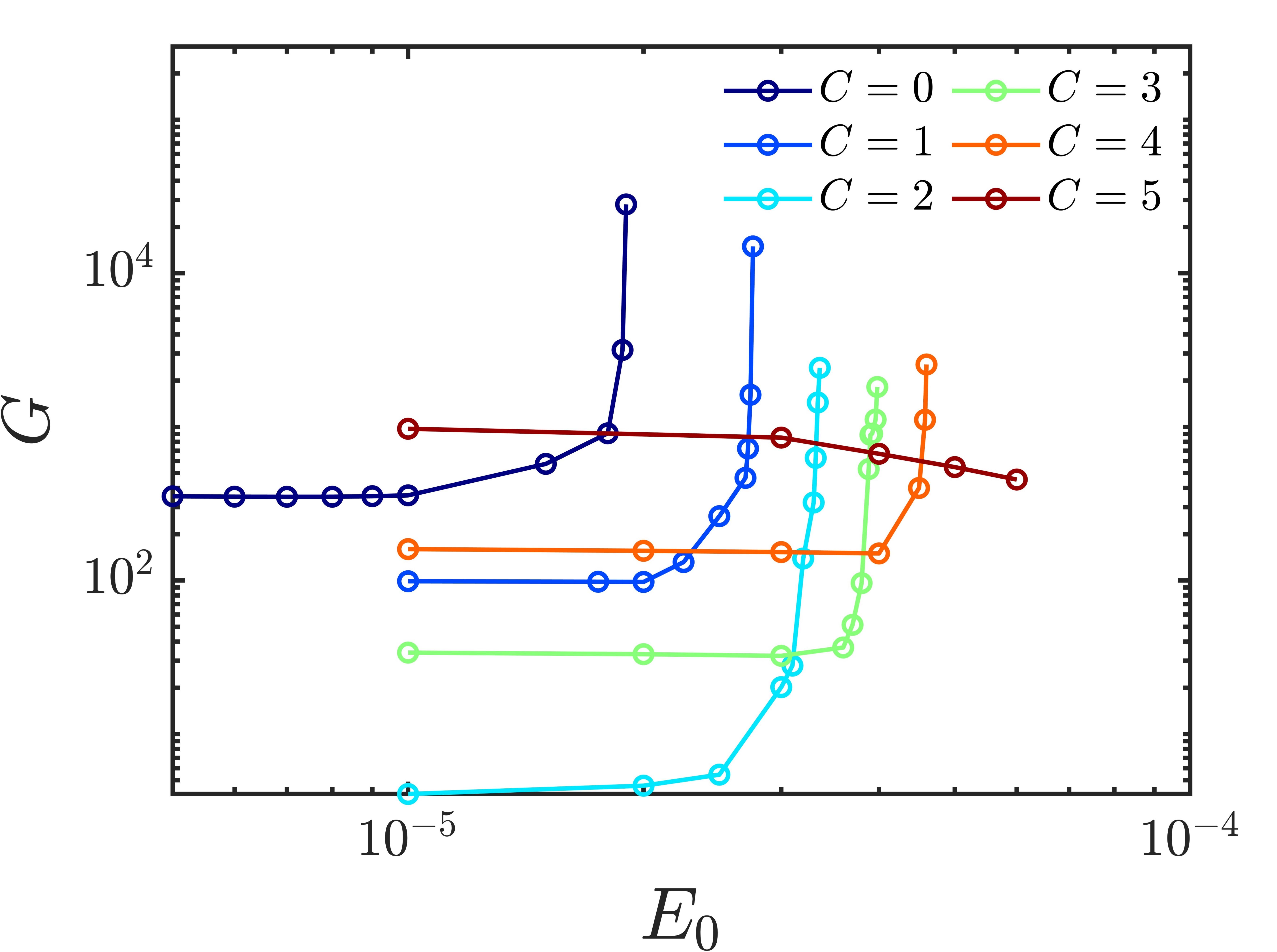}  
          \caption{ \firstref{$(a)$ The residual  $ \delta\mathcal L/\delta \vec{u}(x, 0) $ (left) and energy growth $G=\langle u(\Topt)^2\rangle /\langle u(0)^2\rangle$ (right) versus iterations in the case at $C=1, E_0=2.7e-5, Re=3000$;}  $(b)$ The optimal energy growth $G$ changes with $E_0$ for $C=0-5, \Topt=300, Re=3000$.}     
          \label{G-E0}
        \end{figure}

        To calculate the largest growing perturbation to the laminar state, we use the method of section 2.3. 
        \firstref{Following \cite{pringle2012minimal}, $\Topt$ needs to be sufficiently large to produce a rapid jump in the objective function with 
        respect to a relatively small increase in $E_0$.  
        We first consider a target time of $\Topt=300$, which is approximately
        twice the time at which peak linear growth can be observed at this 
        Reynolds number.
        }
     This $\Topt$ is found to be sufficient to isolate effects of buoyancy on the NLOP. 
    \firstref{Typical residuals $ \delta\mathcal L/\delta \vec{u}(x,0) $ and  energy growth  $G=\langle \vec{u}(\Topt)^2\rangle/\langle \vec{u}(0)^2\rangle$ 
    during the optimisation are shown in figure \ref{G-E0}$(a)$.  A clear drop of residual leads to final optimal energy growth, which means the calculation is well converged, and starting each calculation from different random initial
    velocity fields produces the same final optimal.
    Usually, the calculation is stopped when the change in $G$ is smaller than $10e-5$.}    

     We first 
     present the results at $C=0-5$, where similar NLOPs are found,
     then a new NLOP at $C=6$ is presented separately.
    The optimal energy growth 
    as a function of $E_0$ for several $C$ is plotted in figure \ref{G-E0}($b$). For each $C$ 
   at small $E_0$, the growth $G$ is constant, which implies that the linear optimal has been found.  
   As $E_0$ is increased, the NLOP takes over and $G$ increases.  
   At slightly larger $E_0$ still, a clear jump in the energy growth is observed when shear turbulence is triggered.  Note, however, that at $C=5$ the NLOP is not observed to take over for the given $\Topt=300$.

\begin{figure}
         	\centering
         	\includegraphics[angle=0,width=1\textwidth]{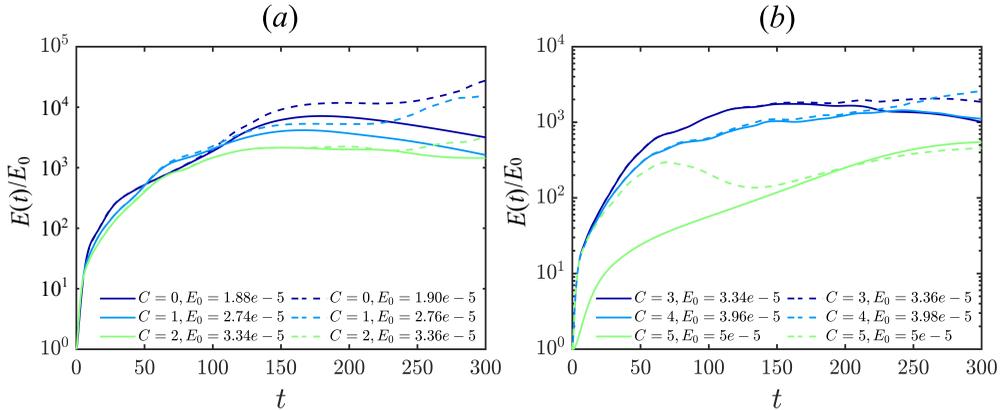}  
         	\caption{The energy growth of NLOP versus time at $(a)$ $C=0-2$ and $(b)$ $C=3-5$. In $(b)$
            \thirdref{for $C=5$, the dashed line is for an NLOP
            of the same type as for $C<5$, while for the solid line, the NLOP is close to the linear eigenfunction.
            }
          }   
         	\label{G-C0-5}
        \end{figure}

\begin{figure}
         	\centering
         	\includegraphics[angle=0,width=1\textwidth]{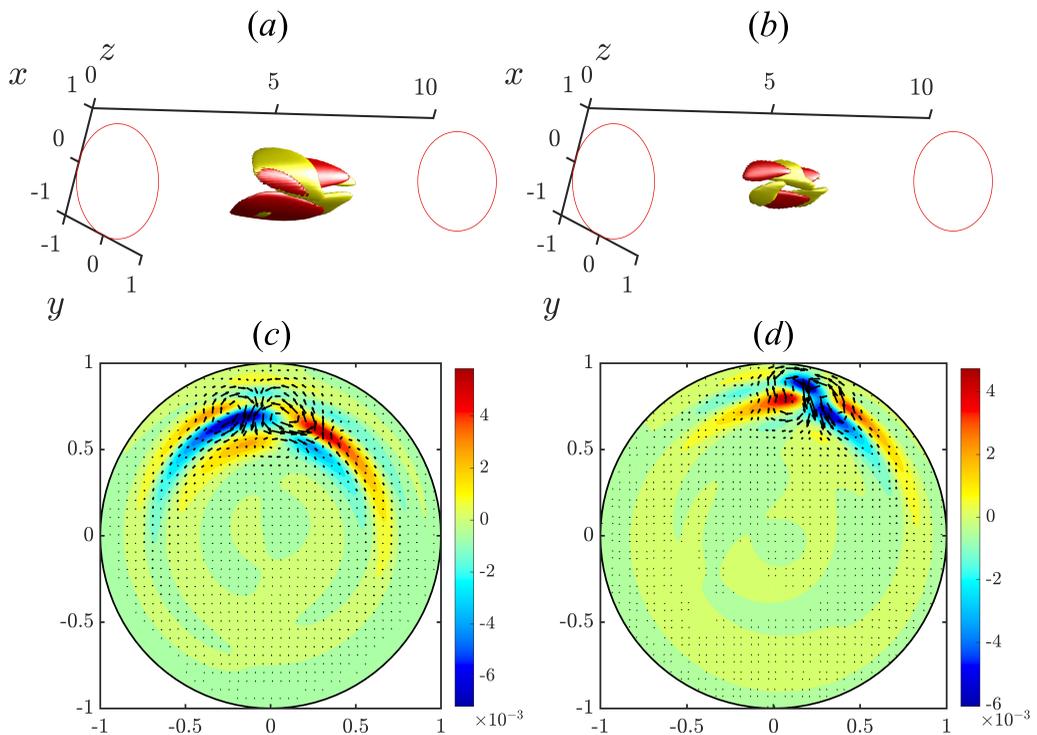}  
         	\caption{Isosurfaces of streamwise velocity perturbation and cross-sections at $z=5$ for the NLOP.
          ($a,c$) $C=0, E_0=1.9e-05$ and ($b,d$) $C=2,E_0=3.36e-05$ at $Re=3000$. Red/yellow isosurfaces are for $30\%$ of the min/max streamwise velocity. Arrows show the cross-stream components; the largest arrow has a magnitude $1.55e-04$ in $(a)$ and $2.39e-4$ in $(b)$.   }         
         	\label{NLOP-C03}
        \end{figure}

    Figure \ref{G-C0-5} shows the time series of the energy growth of an NLOP (solid line) and minimal seed (dash line) for $C=0-5$.   The minimal seed triggers turbulence, while the perturbation eventually decays when it starts from the NLOP at a slightly lower  $E_0$.
    (The true value of $E_0$ for the minimal seed
    could be further refined, but without 
    a noticeable change in the velocity field.
    We will call the perturbation at the upper $E_0$ the minimal seed.)
    At a larger $C$,  the maximum energy growth is smaller and the peak occurs at an earlier time.  The reduced maximum energy is caused by the more flattened base velocity profile, which suppresses the nonlinear instability \citep{marensi2019stabilisation}.  The maximum occurring at an earlier time is attributed to the smaller length scale of the streamwise vortices structure in the minimal seed,  shown figure \ref{NLOP-C03}$(d)$. 
    
    At $C=5$, the algorithm converges to either the solid green line or the dashed green line in figure \ref{G-C0-5}$(b)$, depending on the initial guess, although the solid green case appears more frequently. 
    Note that the two optimals have similar energy
    at the target time $\Topt$, leading to the 
    possibility of local optimals of the Lagrangian. 
    \thirdref{(Local optimals with similar $\mathcal{L}$ have also been observed by \cite{motoki2018optimal} for maximal heat transfer in steady plane Couette flow.)}
    The solid green line includes a period of exponential
    growth, indicating that the perturbation approaches the 
    unstable eigenfunction of the unstable laminar 
    state.  The initial state for the dashed line is of the 
    same type as the NLOP calculated for $C<5$.
        
\begin{figure}
         	\centering
         	\includegraphics[angle=0,width=0.9\textwidth]{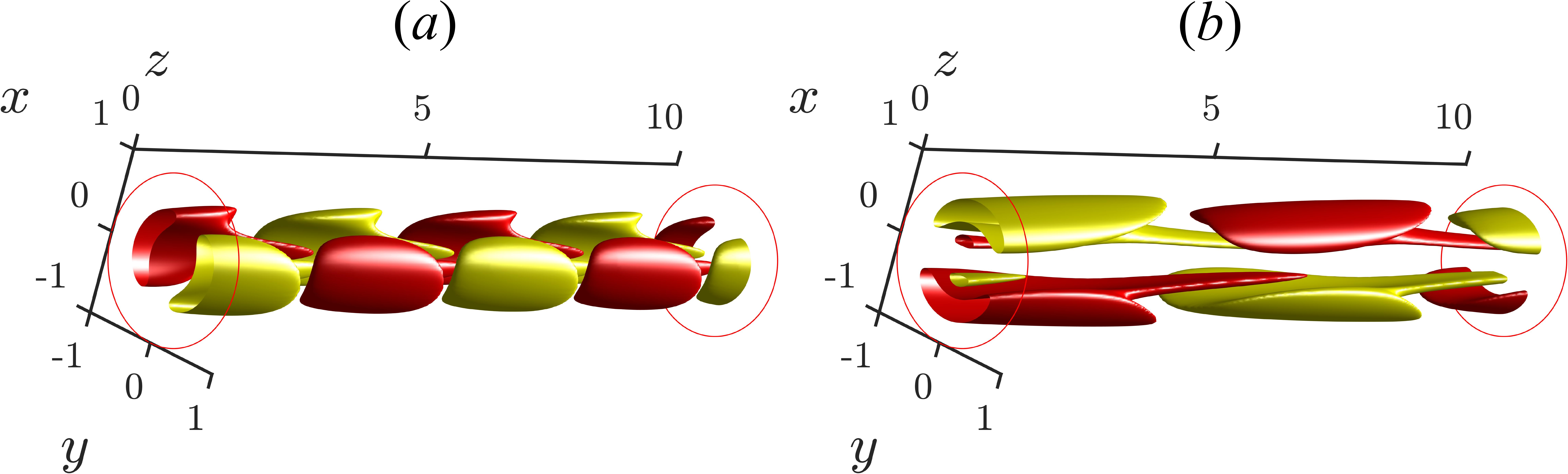}  
         	\caption{ Isosurface of the streamwise velocity of the NLOP at ($a$) $C=4$  and ($b$) $C=5$ at $Re=3000, \Topt=400$, red/yellow are $30\%$ of the min/max value.}         
         	\label{NLOP-C045}
           \end{figure}

    The structures of the minimal seeds at $C=0,2$ are shown in figure \ref{NLOP-C03}. They are essentially similar, localised both in the  azimuthal and streamwise direction, consistent with the isothermal NLOP \citep{pringle2010using,pringle2012minimal}.  But compared with isothermal pipe flow $C=0$, the minimal seed at $C=2$ is more localised and `thinner', being more located closer to the wall, see figure \ref{NLOP-C03}($c,d$).  A thinner minimal seed was also reported by \cite{marensi2019stabilisation} for a (prescribed) flattened base velocity profile. Here the flattened profile is due to the buoyancy.
The effective radial interval for the lift-up mechanism is narrower, and consequently the optimal rolls gradually become thinner.  Structures of the NLOP at $C=3,4$ (not shown)
are similar to figure \ref{NLOP-C03}($d$) but thinner still.
     The optimal perturbation at $C=5$ which is close to the unstable eigenfunction is shown in figure \ref{NLOP-C045}$(b)$, accompanied by the same branch of optimal perturbation at $C=4$ in figure \ref{NLOP-C045}$(a)$,
     found by using a longer target time $\Topt=400$
     (to enable slower linear growth to compete with the NLOP). 
     This type of optimal perturbation is almost unchanged at small $E_0$, suggesting linearity.  Its structures are initially also close to the wall, but after an initial transient growth, they are rolled up near the centre of the pipe, approaching the unstable eigenfunctions. The exponential growth rate is in good agreement with the linear stability analysis.

\begin{figure}
         	\centering
         	\includegraphics[angle=0,width=1\textwidth]{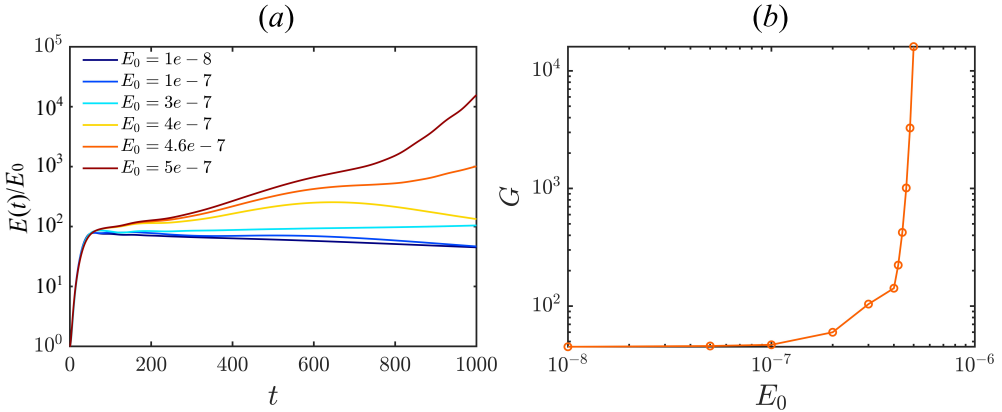}  
         	\caption{Optimal energy growth for
            $C=6,Re=3000,\Topt=1000$.}         
         	\label{G-C06}
    \end{figure}

    For the interesting case at $C=6$, an NLOP with
    structure similar to those found at $C<5$ was not
    found for similar $E_0$ and $\Topt$.
    By reducing the initial energy and extending the target time substantially ($E_0 \sim O(1e-7)$, $\Topt=1000$), we finally located a new type of NLOP.  
    \firstref{(Note that the energy of the convective
    state itself is also much reduced compared to shear 
    turbulence, seen in figure \ref{Flow-states}(a).)}
    Energy growth as a function of time is shown in figure \ref{G-C06}. 
    Optimals at all energies experience a period of energy growth at the beginning. Then, for a small initial energy $E_0=1e-8$, it exponentially decays. At larger initial energy, below the critical energy, the energy grows a little but finally still decays, albeit slowly. For this reason the long optimisation time $\Topt=1000$ was required. At a slightly larger $E_0$ still, the critical energy jump is observed, see figure \ref{G-C06}($b$).

\begin{figure}
         	\centering
         	\includegraphics[angle=0,width=1\textwidth]{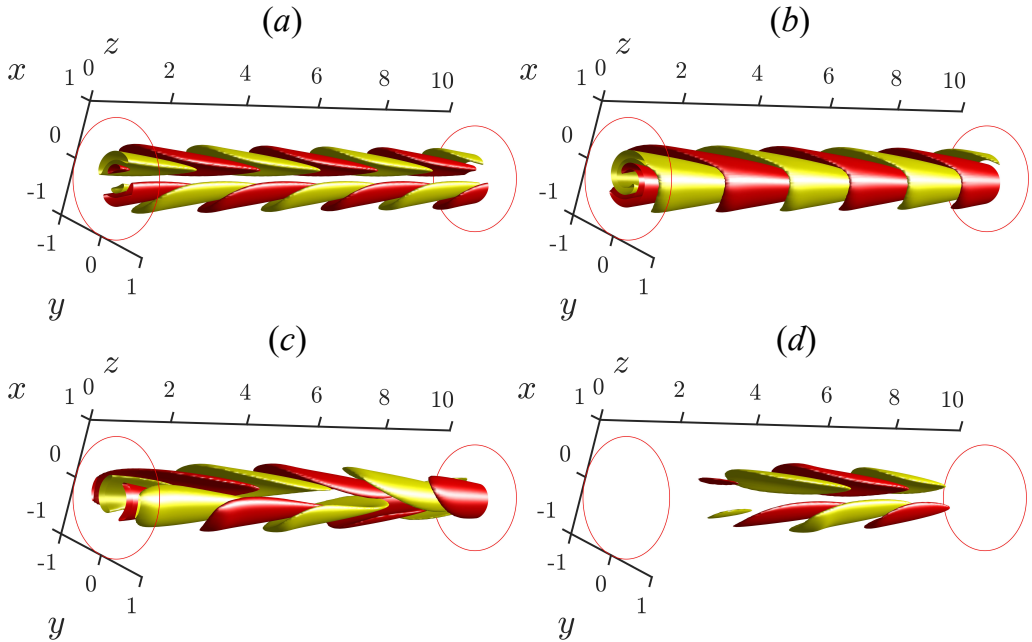}  
         	\caption{Isosurfaces of streamwise perturbation at $C=6$.  Linear optimal ($a$) $E_0=1e-8$, and NLOPs at ($b$) $E_0=1e-7$, ($c$) $E_0=3e-7$, ($d$) $E_0=4e-7$. Red/yellow are for 30$\%$ of the min/max.}         
         	\label{NLOP-C06}
        \end{figure}

\begin{figure}
         	\centering
         	\includegraphics[angle=0,width=0.5\textwidth]{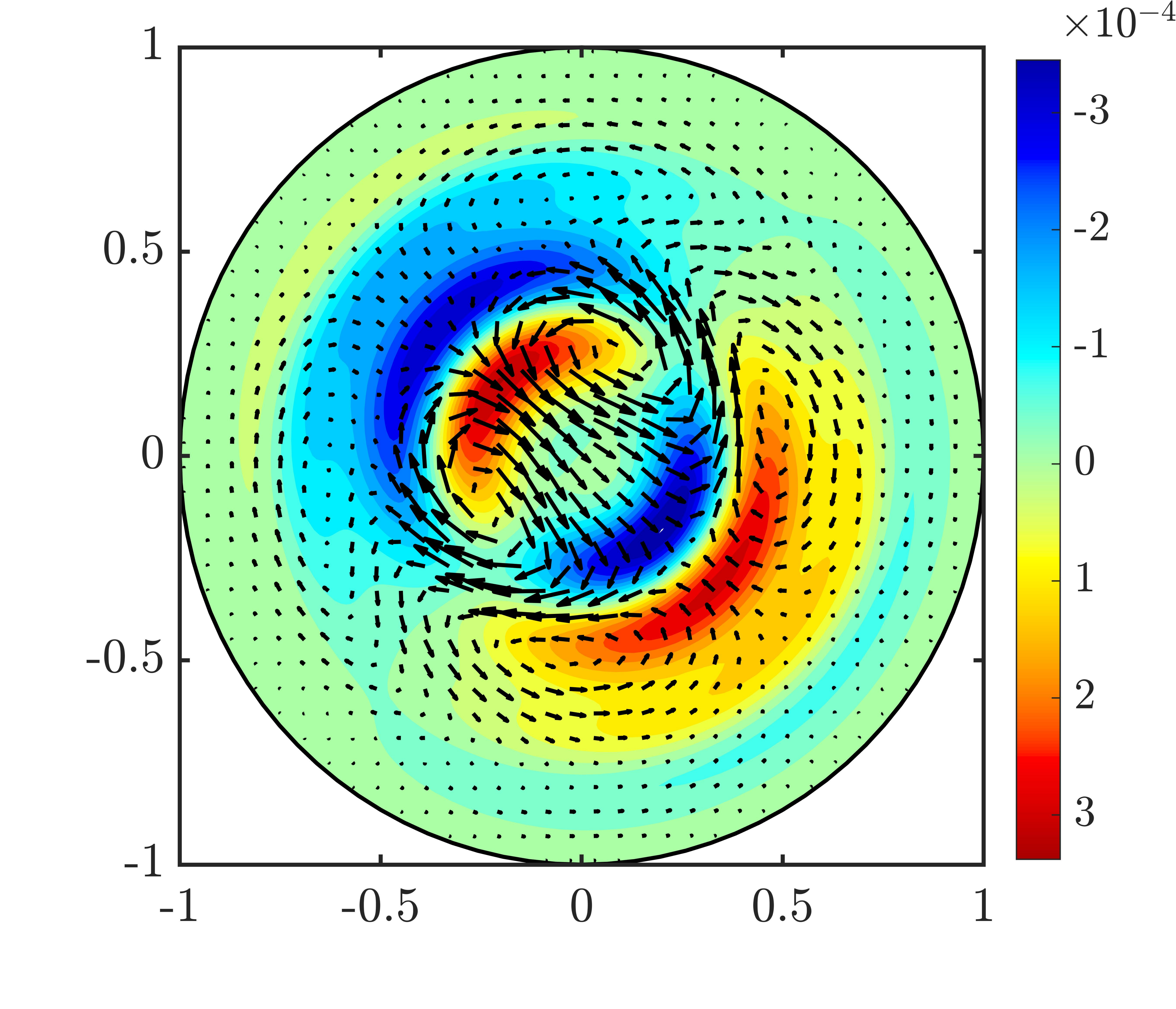}  
         	\caption{Cross-section of NLOP at $C=6$, $E_0=4e-7$.  Contours for streamwise perturbations, while the arrows represent the small cross-stream components; the largest arrow has a magnitude $4.44e-07$.}      
         	\label{Contour-Z5-C06}
        \end{figure}

    Isosurfaces of the optimal at different $E_0$ for 
    $C=6$ are shown in figure \ref{NLOP-C06}.  
    Unlike the NLOP at $C=0-4$, its structure is 
    observed to change as $E_0$ is increased.
    The linear optimal is shown in 
    figure \ref{NLOP-C06}($a$), which is distributed regularly in the streamwise direction.  It is very different from the linear optimal of the isothermal case, which consists of streamwise rolls, and is different from the convective eigenfunctions of figure \ref{unstable-mode}.
    For larger $E_0$, the NLOP shown in \ref{NLOP-C06}($b$), has a more spiral structure.   
    As the initial energy increases further in ($c,d$), it localises in the streamwise direction.
    The last of these, \ref{NLOP-C06}($d$), is the minimal seed which eventually triggers turbulence.  Its cross-section, 
    figure \ref{Contour-Z5-C06}, shows two pairs of very weak streamwise vortices in the centre of the pipe, with very 
    small magnitude compared to the streamwise perturbation. 
    The dynamics of this optimal are investigated in the following 
    section.

\subsection{Transition to convective turbulence}

\begin{figure}
         	\centering
         	\includegraphics[angle=0,width=1\textwidth]{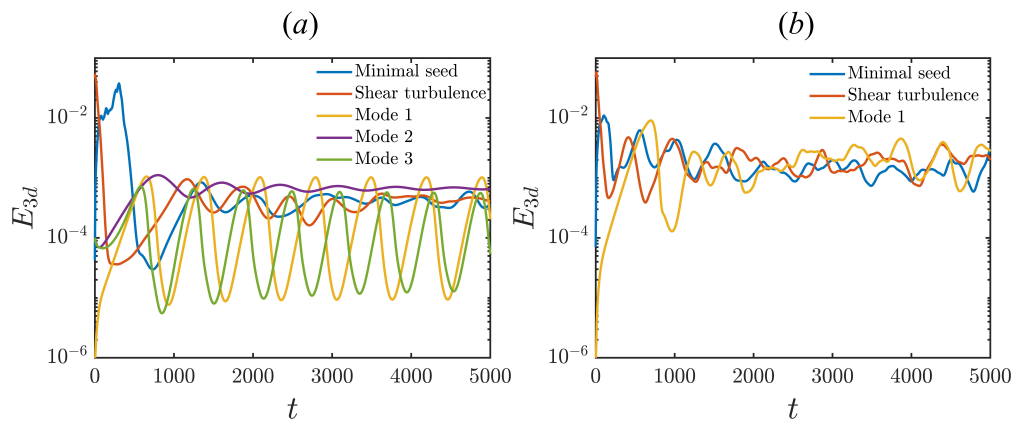} 
         	\caption{$E_{3d}$ 
          (energy of streamwise-dependent component of the flow) for different initial conditions at ($a$) $C=4$ and ($b$) $C=5$ at $Re=3000$. There are three unstable modes at $C=4$: mode 1 ($m=1, k=2$), mode 2 ($m=1, k=1$), mode 3 ($m=1, k=3$). There is only one unstable mode at $C=5$ ($m=1, k=1$).  }         
         	\label{Transition-IC}
    \end{figure}

    Transition to shear turbulence via the NLOP has been widely researched for isothermal flow \citep{pringle2010using,pringle2012minimal,marensi2019stabilisation}, therefore we focus on the transition to convective turbulence ($C\ge 4$) here. 
    In figure \ref{Transition-IC}($a$), transition 
    at $C=4$ and $\Rey=3000$ is 
    compared for several initial conditions --- 
    shear-turbulence, the minimal seed, and unstable eigenfunctions.  $E_{3d}$ is the energy of the 
    streamwise-dependent component of the flow
    (modes $k\ne 0$ in the Fourier expansion (\ref{eq:discr})).
    For the minimal seed, $E_{3d}$ increases 
    substantially at first, reaching the shear-turbulent
    state.  It then decays to the convective turbulence state.  This implies that shear turbulence can still be triggered in the buoyancy regime, but it is not sustained.  
    Starting with a shear turbulent state taken from an isothermal
    simulation, 
    there is an immediate decay at first, then $E_{3d}$ grows exponentially towards convective turbulence. 
    Interestingly, starting from the unstable eigenfunctions, the dynamics approaches a travelling wave solution (mode 2) or relative periodic solutions (mode 1 and mode 3). These states can be approached for a long time, suggesting that the travelling wave and relative periodic orbit solutions are only weakly unstable.
    An equilibrium state and periodic motion have also been reported in early research \citep{kemeny1962combined,scheele1962effect,yao1987fully}, but at low Reynolds number $(Re<2000)$.

    At $C=5$, figure \ref{Transition-IC}($b$), the unstable eigenfunction does not lead to a travelling wave or periodic solution, and instead transitions to convective turbulence directly.
    This may be due to the enhanced instability of these equilibrium solutions at larger $C$. 
    The minimal seed at $C=5$ also
    triggers the convective turbulence directly, 
    or at least, there is no period of decay after 
    the initial growth, which would indicate a distinct
    switch from  shear turbulence to convective
    turbulence. Meanwhile, starting from shear turbulence, there is a clear
    initial decay, followed by exponential growth at a rate which matches that starting from the eigenfunction.

\begin{figure}
         	\centering
         	\includegraphics[angle=0,width=0.6\textwidth]{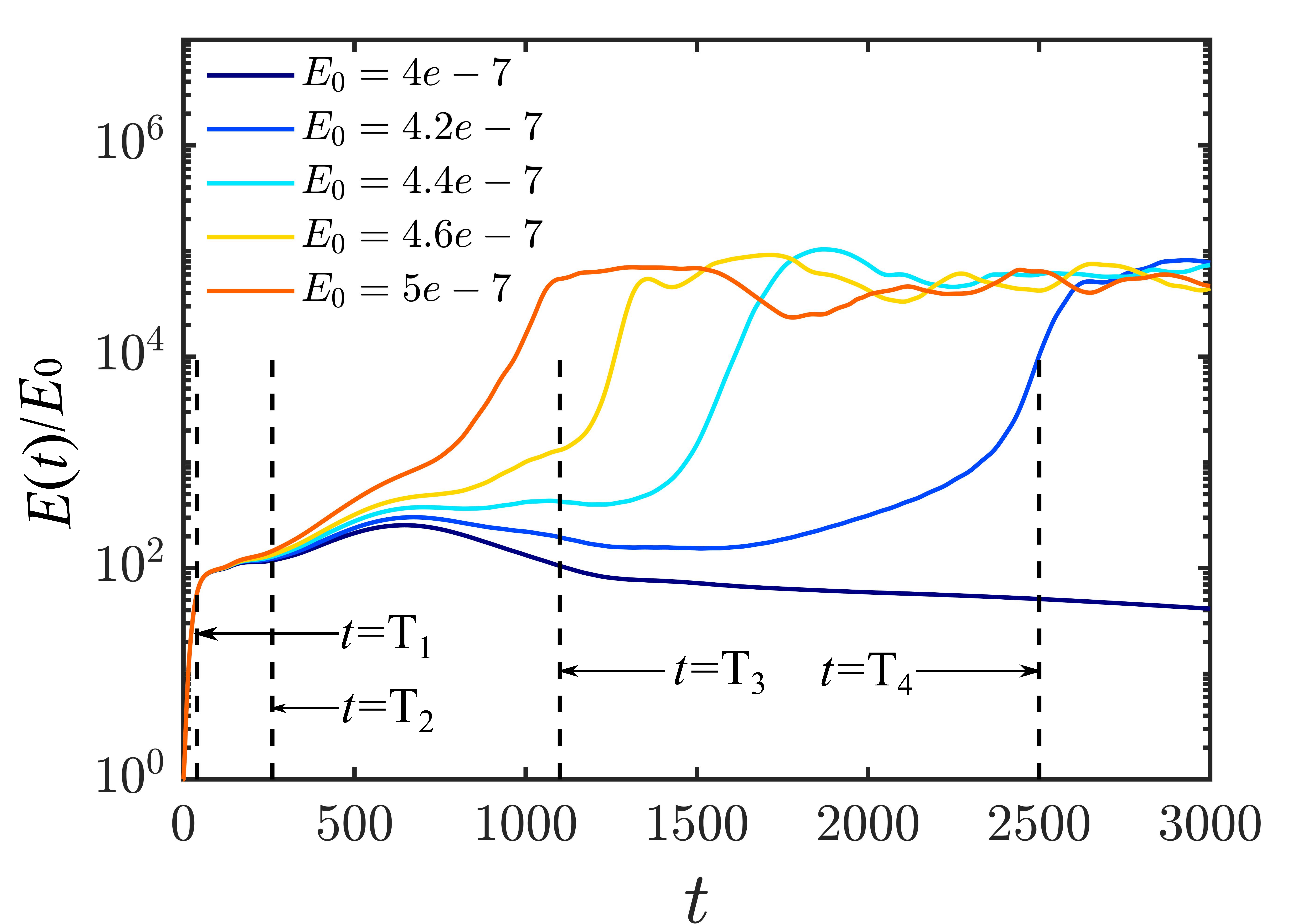} 
         	\caption{Time series of energy starting from the NLOP obtained at several $E_0$, at $C=6, Re=3000$. 
          The cases $E_0=4e-7, 4.6e-7, 5e-7$ were also shown in figure \ref{G-C06}($a$) up to $t=\Topt=1000$. }         
         	\label{Transition-MSC06}
        \end{figure}

\begin{figure}
         	\centering
         	\includegraphics[angle=0,width=1\textwidth]{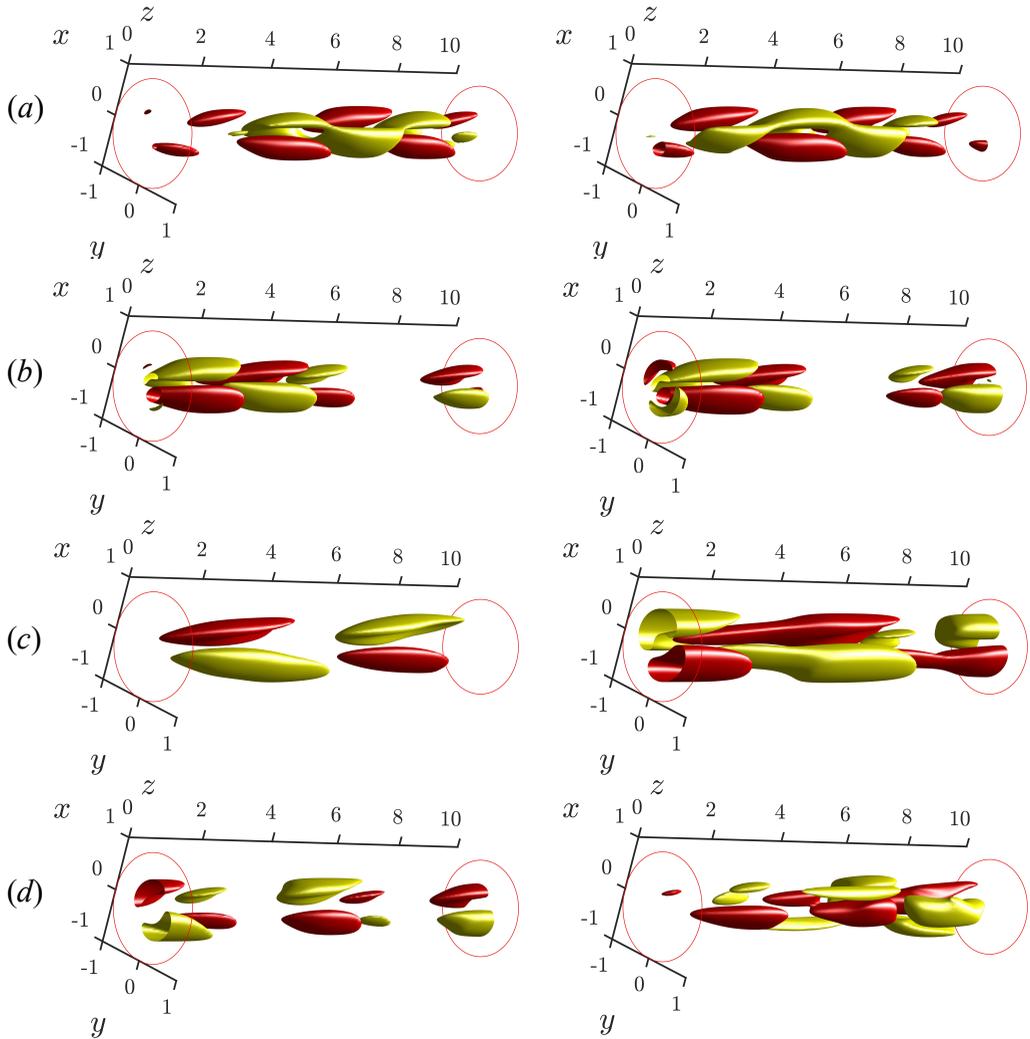} 
         	\caption{Isosurfaces of streamwise velocity 
          starting from NLOPs for ({\it left}) $E_0=4e-7$ and ({\it right})
          $E_0=4.2e-7$           
          at times ($a$) $t=T_1$, ($b$) $t=T_2$, ($c$) $t=T_3$, ($d$) $t=T_4$.
          $C=6, Re=3000$.  Times are marked in figure \ref{Transition-MSC06}.  Red/yellow are at 30$\%$ of the min/max value.}  
         	\label{Transition-MSC06-iso}
        \end{figure}

    The transition process at $C=6$, 
    where the laminar state is linearly stable, is quite different,
    shown in
    figure \ref{Transition-MSC06}.
    Compared with transition via the NLOP in the isothermal case 
    \citep{pringle2012minimal}, the transition process is much slower:
    the earliest time turbulence is seen at approximately $t=1000$
    in the figure \ref{Transition-MSC06}, and hence the large target time $\Topt=1000$ required.
    The edge state \citep{duguet2008transition}, here to the convective turbulent state, appears to be 
    less unstable than that for the shear turbulent state,
    as intermediate energies before transition can be achieved for long
    times with little refinement of the initial energy.
    Some typical coherent structures that appear in the process of transition are presented in figure \ref{Transition-MSC06-iso},
    starting from slightly different initial energies,
    $E_0=4e-7$ ({\it left}) and $E_0=4.2e-7$ ({\it right}). Initially, the contours of minimal seed are tightly layered and forward-facing, but because the base profile is M-shaped, the structures are inclined into the shear, see figure \ref{NLOP-C06}$(d)$.  (For the 
    isothermal minimal seed, similar layers are inclined into the shear, but without the region of reversed flow, they are 
    backward-facing and located closer to the wall \citep{pringle2012minimal}.)  Therefore, between $t=0$ and $t=T_1$ there is a great energy growth in a short time through the Orr mechanism \citep{orr1907stability,pringle2012minimal}.
    The flow then evolves to a simpler organised state at $t=T_2$, figure \ref{Transition-MSC06-iso}$(\it b)$, and both flows still look similar.  At $t=T_3$ $(\it c, right)$, the structure is more elongated, and looks similar to the travelling wave solution of figure \ref{TW}$(a)$.
    For slightly lower initial energy, 
    the state gradually decays exponentially $T_4$ ({\it d, left}) with a very small decay rate.    
    For the slightly larger initial energy, it evolves into the convective state at $T_4$ ({\it d, right}).  
   A relatively rapid increase of energy is observed around $T_4$
   for this case ($E_0=4.2e-7$), and a similar final growth stage is observed for the NLOPs of larger initial energies, at earlier times, seen in figure \ref{Transition-MSC06}. 
   Overall, the transition process of minimal seed is complex and long compared to that of the minimal seed for transition to shear-turbulence.

    \subsection{Travelling wave and periodic orbit solutions}

    In recent years, a series of invariant solutions of the Navier--Stokes equations have been found for pipe flow \citep{wedin2004exact,hof2004experimental,eckhardt2007turbulence} and other shear flows \citep{nagata1990three,toh2003periodic,reetz2019exact}. These invariant solutions have significantly enriched our understanding of the transition and maintenance of turbulence. \thirdref{A subset of these invariant solutions are found in the edge between the laminar and turbulent attractors, and shed some light on the transition to turbulence \citep{duguet2008transition, avila2023transition}. Within turbulence, a well-known periodic solution for plane Couette flow \citep{kawahara2001periodic} successfully reproduces the self-sustaining process of shear turbulence 
    \citep{hamilton1995regeneration}. }  
    Although these solutions are unstable, the dimensions of their unstable manifolds in phase space are typically low \citep{kawahara2005laminarization,kerswell2007recurrence}.  A generic turbulent state may approach them and spend a substantial fraction of its lifetime in their neighbourhood \citep{kawahara2012significance}.

\begin{figure}
         	\centering
         	\includegraphics[angle=0,width=1\textwidth]{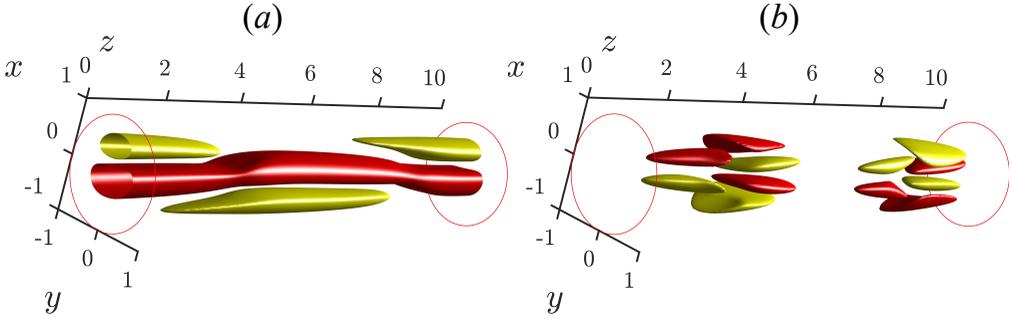}  
         	\caption{Travelling wave solution approached during transition at $C=4, Re=3000$.  Isosurfaces of $(a)$ streamwise perturbation and $(b)$ streamwise vorticity. Red/yellow are at 30$\%$ of the min/max value.}   
         	\label{TW}
        \end{figure}

    In figure \ref{Transition-IC}, the dynamics approached a travelling wave solution and periodic solutions. 
    Using states from the trajectories seen in
    figure \ref{Transition-IC}(a) at later times
    as initial guesses, 
    the Jacobian-free Newton–Krylov (JFNK) method 
    \citep{willis2019equilibria} converged very quickly to a travelling wave solution and periodic solutions. 
    The travelling wave solution is plotted in figure \ref{TW}, which has a wave speed $c=0.6783$.
    Streamwise vorticity is localised in streamwise direction, located at the bends of the high-speed streak.  
    \thirdref{
    An initial state was also taken from $t=1500$ for $E_0=4.2e-7$
    of figure \ref{Transition-MSC06}, and rapidly converged to a
    periodic orbit with weak variation and period $68.4$.    
    Linear instability analysis of this periodic orbit solution shows that it has only one real eigenvalue
     with small real part, $\Re(\sigma)=8.06e-4$, 
    so that excessive refinement of the initial energy is not required to stay close to it for significant time
    before diverging to the convective turbulence or the laminar state (stable at $C=6$).  
    As this periodic orbit has only one unstable direction, it is an attractor within the boundary between
    the laminar and convective states, i.e.\ it is an `edge state'  
    \citep{duguet2008transition}. }

\begin{figure}
         	\centering
         	\includegraphics[angle=0,width=1\textwidth]{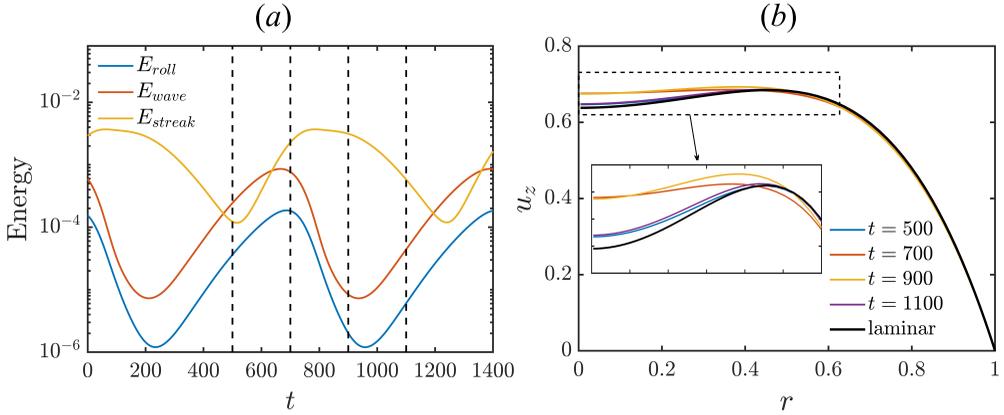}  
         	\caption{$(a)$ Time evolution of the 
          periodic solution $PO_1$ at $C=4,Re=3000$, where $E_{roll}=E(u_r)+E(u_\phi)$, $E_{wave}=E(u_z)_{k\ne 0}$, $E_{streak}=E(u_z)_{k=0}$. $(b)$ the turbulent mean velocity profile at several \thirdref{times}}   
         	\label{E_PO}
        \end{figure}

\begin{figure}
         	\centering
         	\includegraphics[angle=0,width=1\textwidth]{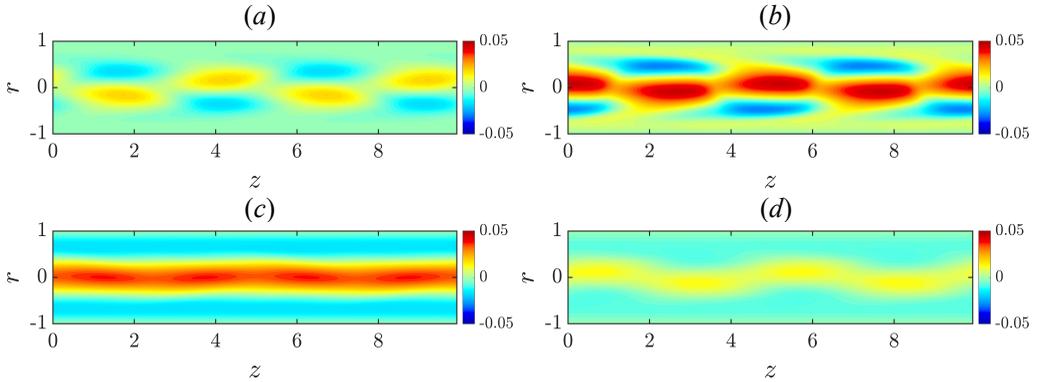}  
         	\caption{The contour of periodic solution $PO_1$ at $(a) t=500$,$(b) t=700$,$(c) t=900$,$(d) t=1100$, colored by streamwise velocity disturbance 
          $u$.}     
         	\label{Uz_PO}
        \end{figure}

    The periodic solutions approached from the first and third unstable eigenfunctions at $C=4$, marked $PO_1$ and $PO_2$ respectively, have long periods, $T_1=724.4$ and $T_2=623.72 $. They are similar, only with a different dominant axial wave number ($k=2$ for $PO_1$ and $k=3$ for $PO_2$).  We here pay attention to the periodic solution triggered by the first unstable mode.  Time series of energies \firstref{over approximately two periods of $PO_1$} are shown in figure \ref{E_PO}$(a)$, the energy of the rolls, $E_{roll}=E(u_r)+E(u_\phi)$,
    streaks, $E_{streak}=E(u_z)_{k=0}$, and waves,
    $E_{waves}=E(u_z)_{k\ne 0}$, where the subscripts $k=0$ and $k\ne0$ indicate streamwise independent and dependent components extracted via the Fourier decomposition.
    As a single streak forms at the centre of the pipe, we have not
    split azimuthal dependence through $m$.
    For the well-known periodic process studied in channel flow
    by \cite{hamilton1995regeneration},
    the peak of the roll energy closely corresponds to the valley of the streak energy, 
    but for $PO_1$, these
    energies are more offset.
    Contours of streamwise perturbation velocity $u_z$ at $t=500,700,900,1100$ are plotted in figure \ref{Uz_PO}, which shows how the structure changes over one period. 
    At $t=500$, the streak energy is at its lowest, and the velocity field resembles the unstable eigenfunction, only with a weak break of shift and reflect symmetry.  Then, by $t=700$ the unstable eigenfunction has grown until regions of positive velocity have merged, forming a high-speed streak
    in the centre of the pipe.  
    At this stage, the wavy high-speed streak in the centre of the pipe is similar to that of the travelling wave of figure \ref{TW}.
    In the
    presence of the streak, the roll perturbations
    decay, $t=900$, then return once the streak has weakened at $t=1100$.
    Therefore, three typical stages are observed in the periodic solution, i.e.\ growth of the eigenfunction, formation of the streak and decay of the streak. 
    These processes can be understood from the changes in the mean velocity profile, seen in figure \ref{E_PO}$(b)$. At first, when the perturbation is weak, the velocity profile is  close to the laminar solution $(t=500)$. Linear instability takes charge of the dynamics, and the unstable eigenfunction grows at this stage. Then, when the perturbation excites a sufficiently strong streak near the axis, the velocity profile is flattened $(t=700,900)$ and the linear instability is suppressed, causing the perturbation to decay. Once the perturbation decays to a certain level, the velocity profile becomes unstable again $(t=1100)$ and the process repeats.  (The full periodic process is shown in the supplementary movie 1.) 
    
\begin{figure}
         	\centering
         	\includegraphics[angle=0,width=1\textwidth]{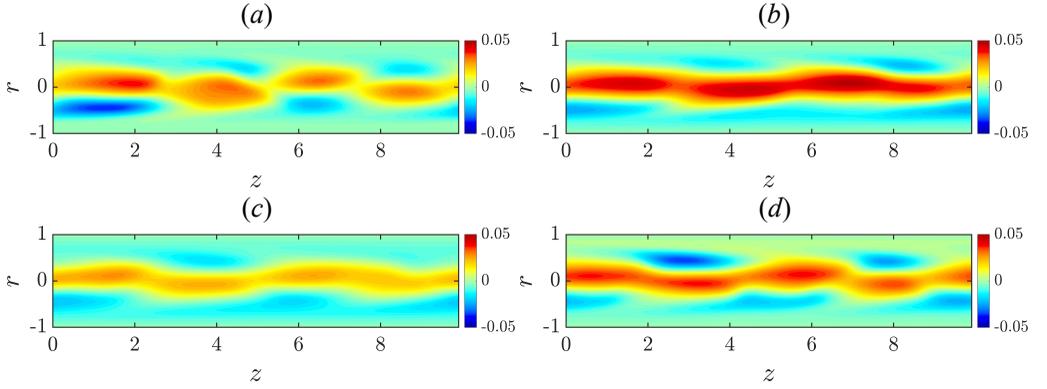}  
         	\caption{The contour of transient convective turbulence at $(a)$ $t=2500$, $(b)$ $t=2750$, $(c)$ $t=3000$, $(d)$ $t=3250$, picked from figure \ref{Transition-IC} when the turbulence is triggered by shear turbulence at $C=4$, the contour is coloured by streamwise velocity disturbance $u$.}     
         	\label{Uz_arbi}
        \end{figure}

    
    In the chaotic convective state at $C=4$, 
    similar phenomena to that of the periodic orbits occurs, 
    but the laminar flow is unstable to multiple modes.  Contours of convective turbulence show that the flow 
    still exhibits the three typical processes, i.e, the growth of unstable modes (figure \ref{Uz_arbi}$(a)$), the formation of the high-speed streak (figure \ref{Uz_arbi}$(b)$), decay of the streak and reappearance of the unstable modes
    (figure \ref{Uz_arbi}$(c,d)$).  
    The scales of the observed modes are in good agreement with the first unstable mode and the third unstable mode ($k=2$ and $k=3$), which is consistent with the flow wandering the two periodic solutions. 
    The scale of the travelling wave ($k=1$)
    is hardly observed. 
    
\begin{figure}
         	\centering
         	\includegraphics[angle=0,width=1\textwidth]{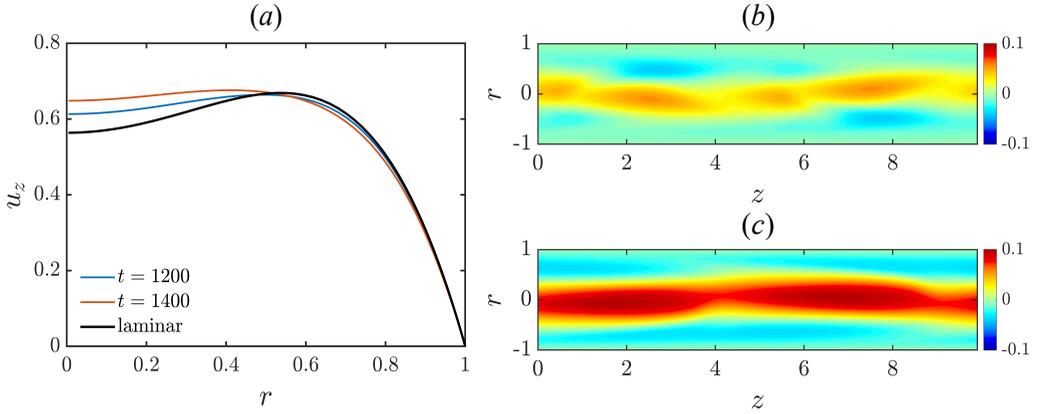}  
         	\caption{$(a)$ The turbulent mean profiles of the flow at $(b)$ and $(c)$. The contour of transient convective turbulence at $(b)$ $t=1150$, $(c)$ $t=1350$, picked from figure \ref{Transition-IC} when the turbulence is triggered by shear turbulence at $C=5$, the contour is colored by velocity disturbance $u$.}    
         	\label{Uz_c05}
        \end{figure}

    At larger $C$ the periodic solutions become more unstable and the flow is more chaotic. However, the typical process can be still observed, as in figure \ref{Uz_c05}, which shows two transient flow states taken from the flow at $C=5$. Figure \ref{Uz_c05}$(b)$ shows the typical growth of an unstable mode upon a mean velocity profile of figure \ref{Uz_c05}$(a)$. 
    It is observed that the profile is close to the laminar profile, but does not go so close as the case at $C=4$ of figure \ref{E_PO}$(b)$, as the 
    laminar solution is also more unstable.
    Figure \ref{Uz_c05}$(c)$ shows the high-speed streak formed by growth of the unstable mode, whose profile is almost flattened and loses  linear instability. These typical flow states can be often observed at $C=5$, and can even be observed at $C=10$, although less clearly.
    Therefore, it appears that the convective turbulence is self-sustained by these typical stages, i.e.\ the growth of unstable modes, the formation of streaks and the decay of streaks.
        
\section{Conclusion}
  \label{sec:headings} 
  In this study, a new DNS model that includes 
  time-dependent heat flux and background temperature
  gradient is established. The results show good consistency with the  experiment and a little improvement 
  over the model of \cite{marensi2021suppression},
  which also had time-dependent heat flux, but a spatially uniform 
  heat sink term.  Simulations at different values of the buoyancy parameter $C$ confirm three typical flow states, i.e.\ shear turbulence, laminarisation and convective turbulence. The distribution of these states in the parameter space  is consistent with the calculations of \cite{marensi2021suppression}. Detailed examination of the convective turbulence found that it lacks near-wall rolls, which enhance the mixing of fluid near the wall and centre, and therefore the heat transfer is seriously reduced. The linear stability analysis further verifies that the convective turbulence is driven by linear instability \citep{yao1987fully,su2000linear,khandelwal2015weakly,marensi2021suppression}. The structures of unstable eigenfunctions are found to be mainly concentrated on the centre of the pipe, which explains the similar flow characteristics observed in the transient convective turbulence.

  Nonlinear non-modal stability analysis is extended to the heated pipe case to determine the effects of buoyancy on the smallest perturbuation that 
  triggers transition, the 
  `minimal seed'.  As $C$ is increased from $0$,
  it is found that the structure of the minimal seed becomes thinner, rolls move nearer to the wall, and the critical energy for transition increases. Maximum energy growth occurs earlier and is greatly reduced.  
  Importantly, this suggests that strategies for exciting shear
  turbulence to encourage greater heat flux, do not
  straightforwardly carry over from the isothermal to the heated pipe case.
  The branch of NLOP that triggers turbulence in isothermal flow
  could not be tracked beyond $C=5$,
  due to the combined reduction in growth and enhancement of linear instability.  
  Instead, a new type of NLOP arises for a longer target time and a small initial energy. The new NLOP changes structure as the initial energy is increased, localising  in the streamwise direction.  It is concentrated towards the centre of the pipe, triggering convective turbulence directly, but over a long  period.

  The transitions towards convective turbulence at $C=4,5,6$ triggered by the NLOP were compared with transition via the unstable eigenfunctions.  
  At $C=4$, it is found 
  that starting from the unstable eigenfunctions leads to a travelling wave solution or periodic solutions, while the minimal seed first triggers shear turbulence, and then the shear turbulence decays to convective turbulence. At $C=5$, both unstable eigenfunctions and the minimal seed trigger convective turbulence directly.  At $C=6$, the transition starting from the new NLOP is much slower. 
  The  edge state to the convective turbulence appears to be less unstable than that of  shear turbulence, and intermediate energies before transition can be achieved for a long time with little refinement of the initial energy.
The approached travelling wave solution and periodic solutions were calculated using Jacobian-free Newton–Krylov (JFNK) method, \thirdref{and their stability
calculated using Arnoldi iteration.}
The periodic orbit solution at $C=6$ has only one unstable \thirdref{direction}, as expected for an edge state.

The periodic solution is found to distill three typical processes, i.e. the growth of an unstable eigenfunction, the formation of streaks and the decay of the streaks.  Analysis of the mean velocity profile showed that the periodic process is caused by the appearance and suppression of linear instability of the mean velocity profile.  
This is fundamentally different from the 
self-sustaining process of isothermal shear flow,
which occurs in the absence of linear instability
of the mean flow.
Chaotic convective turbulence at $C=4$ is still dominated by the three typical processes, but consists of the scales of two periodic solutions,
consistent with the dynamics wandering between the two solutions in phase space. A similar phenomenon is still often observed at $C=5$, and can be recognised even up to $C=10$.  

\firstref{We caution that of our work has focused on the case $Re=3000$.  However, the form of the minimal seed is known to be robust 
in the isothermal case
with respect to large changes in the $\Rey$ and flattening of the base profile \citep{marensi2019stabilisation}.  Due to the collapse with respect to the buoyancy parameter $Bo$ for the transition seen in figure \ref{SCL-boundary}(b), we expect similar behaviour to be exhibited at larger $Re$, in particular, for values of $C$ around where transition from shear turbulence to the convective state is observed.}

\backsection[Supplementary movies]{\label{SupMat}Supplementary movies are available}

\backsection[Acknowledgements]{ This work used the Cirrus UK National Tier-2 HPC Service at EPCC (http://www.cirrus.ac.uk) funded by the University of Edinburgh and EPSRC (EP/P020267/1).} 

\backsection[Funding]{ S.C. acknowledges the funding from Sheffield–China Scholarships Council PhD Scholarship Programme (CSC no. 202106260029).}

\backsection[Declaration of interests]{The authors report no conflict of interest.}


\backsection[Author ORCIDs]
{Shijun Chu, https://orcid.org/0000-0002-3037-6370;
 Ashley P. Willis, https://orcid.org/0000-0002-2693-2952;
 Elena Marensi, https://orcid.org/0000-0001-7173-4923.}





\bibliographystyle{jfm}
\bibliography{jfm2esam}

\begin{thebibliography}{60}
\expandafter\ifx\csname natexlab\endcsname\relax\def\natexlab#1{#1}\fi
\def\au#1{#1} \def\ed#1{#1} \def\yr#1{#1}\def\at#1{#1}\def\jt#1{\textit{#1}}
  \def\bt#1{#1}\def\bvol#1{\textbf{#1}} \def\vol#1{#1} \def\pg#1{#1}
  \def\publ#1{#1}\def\arxiv#1{#1}\def\org#1{#1}\def\st#1{\textit{#1}}

\bibitem[Ackerman(1970)]{ackerman1970pseudoboiling}
{\sc \au{Ackerman, JW}} \yr{1970}  \at{Pseudoboiling heat transfer to
  supercritical pressure water in smooth and ribbed tubes} .

\bibitem[Avila {\em et~al.\/}(2011)Avila, Moxey, De~Lozar, Avila, Barkley \&
  Hof]{avila2011onset}
{\sc \au{Avila, Kerstin}, \au{Moxey, David}, \au{De~Lozar, Alberto}, \au{Avila,
  Marc}, \au{Barkley, Dwight} \& \au{Hof, Bj{\"o}rn}} \yr{2011}  \at{The onset
  of turbulence in pipe flow}.  \jt{Science}  \bvol{333}~(6039),
  \pg{192--196}.

\bibitem[Avila {\em et~al.\/}(2023)Avila, Barkley \& Hof]{avila2023transition}
{\sc \au{Avila, Marc}, \au{Barkley, Dwight} \& \au{Hof, Bj{\"o}rn}} \yr{2023}
  \at{Transition to turbulence in pipe flow}.  \jt{Annual Review of Fluid
  Mechanics}  \bvol{55},  \pg{575--602}.

\bibitem[Bae {\em et~al.\/}(2005)Bae, Yoo \& Choi]{bae2005direct}
{\sc \au{Bae, Joong~Hun}, \au{Yoo, Jung~Yul} \& \au{Choi, Haecheon}} \yr{2005}
  \at{Direct numerical simulation of turbulent supercritical flows with heat
  transfer}.  \jt{Physics of fluids}  \bvol{17}~(10).

\bibitem[Bae {\em et~al.\/}(2006)Bae, Yoo, Choi \& McEligot]{bae2006effects}
{\sc \au{Bae, Joong~Hun}, \au{Yoo, Jung~Yul}, \au{Choi, Haecheon} \&
  \au{McEligot, Donald~M}} \yr{2006}  \at{Effects of large density variation on
  strongly heated internal air flows}.  \jt{Physics of Fluids}  \bvol{18}~(7).

\bibitem[Celataa {\em et~al.\/}(1998)Celataa, Dannibale, Chiaradia \&
  Cumo]{celataa1998upflow}
{\sc \au{Celataa, Gian~Piero}, \au{Dannibale, Francesco}, \au{Chiaradia,
  Andrea} \& \au{Cumo, Maurizio}} \yr{1998}  \at{Upflow turbulent mixed
  convection heat transfer in vertical pipes}.  \jt{International journal of
  heat and mass transfer}  \bvol{41}~(24),  \pg{4037--4054}.

\bibitem[Chen \& Chung(1996)]{chen1996linear}
{\sc \au{Chen, Yen-Cho} \& \au{Chung, JN}} \yr{1996}  \at{The linear stability
  of mixed convection in a vertical channel flow}.  \jt{Journal of Fluid
  Mechanics}  \bvol{325},  \pg{29--51}.

\bibitem[Chen \& Chung(2002)]{chen2002direct}
{\sc \au{Chen, Yen-Cho} \& \au{Chung, JN}} \yr{2002}  \at{A direct numerical
  simulation of k-and h-type flow transition in a heated vertical channel}.
  \jt{Physics of Fluids}  \bvol{14}~(9),  \pg{3327--3346}.

\bibitem[Cherubini \& De~Palma(2013)]{cherubini2013nonlinear}
{\sc \au{Cherubini, Stefania} \& \au{De~Palma, Pietro}} \yr{2013}
  \at{Nonlinear optimal perturbations in a couette flow: bursting and
  transition}.  \jt{Journal of Fluid Mechanics}  \bvol{716},  \pg{251--279}.

\bibitem[Cherubini {\em et~al.\/}(2011)Cherubini, De~Palma, Robinet \&
  Bottaro]{cherubini2011minimal}
{\sc \au{Cherubini, Stefania}, \au{De~Palma, Pietro}, \au{Robinet, J-C} \&
  \au{Bottaro, Alessandro}} \yr{2011}  \at{The minimal seed of turbulent
  transition in the boundary layer}.  \jt{Journal of Fluid Mechanics}
  \bvol{689},  \pg{221--253}.

\bibitem[Chu {\em et~al.\/}(2024)Chu, Willis \& Marensi]{CSJ24}
{\sc \au{Chu, Shijun}, \au{Willis, Ashley~P.} \& \au{Marensi, Elena}} \yr{2024}
  Linear and nonlinear optimisation for laminarisation in pipe flow. In
  preparation.

\bibitem[Chu {\em et~al.\/}(2016)Chu, Laurien \& McEligot]{chu2016direct}
{\sc \au{Chu, Xu}, \au{Laurien, Eckart} \& \au{McEligot, Donald~M}} \yr{2016}
  \at{Direct numerical simulation of strongly heated air flow in a vertical
  pipe}.  \jt{International Journal of Heat and Mass Transfer}  \bvol{101},
  \pg{1163--1176}.

\bibitem[Duguet {\em et~al.\/}(2008)Duguet, Willis \&
  Kerswell]{duguet2008transition}
{\sc \au{Duguet, Yohann}, \au{Willis, Ashley~P} \& \au{Kerswell, Rich~R}}
  \yr{2008}  \at{Transition in pipe flow: the saddle structure on the boundary
  of turbulence}.  \jt{Journal of Fluid Mechanics}  \bvol{613},  \pg{255--274}.

\bibitem[Eckhardt {\em et~al.\/}(2007)Eckhardt, Schneider, Hof \&
  Westerweel]{eckhardt2007turbulence}
{\sc \au{Eckhardt, Bruno}, \au{Schneider, Tobias~M}, \au{Hof, Bjorn} \&
  \au{Westerweel, Jerry}} \yr{2007}  \at{Turbulence transition in pipe flow}.
  \jt{Annu. Rev. Fluid Mech.}  \bvol{39},  \pg{447--468}.

\bibitem[Guo {\em et~al.\/}(2007)Guo, Zhu \& Liang]{guo2007entransy}
{\sc \au{Guo, Zeng-Yuan}, \au{Zhu, Hong-Ye} \& \au{Liang, Xin-Gang}} \yr{2007}
  \at{Entransy—a physical quantity describing heat transfer ability}.
  \jt{International Journal of Heat and Mass Transfer}  \bvol{50}~(13-14),
  \pg{2545--2556}.

\bibitem[Hall \& Jackson(1969)]{hall1969laminarization}
{\sc \au{Hall, William~Bateman} \& \au{Jackson, JD}} \yr{1969} Laminarization
  of a turbulent pipe flow by buoyancy forces.  \bt{In {\em Mechanical
  Engineering\/}}, ,  \vol{vol.~91},  \pg{p.~66}. ASME-AMER SOC MECHANICAL ENG
  345 E 47TH ST, NEW YORK, NY 10017.

\bibitem[Hamilton {\em et~al.\/}(1995)Hamilton, Kim \&
  Waleffe]{hamilton1995regeneration}
{\sc \au{Hamilton, James~M}, \au{Kim, John} \& \au{Waleffe, Fabian}} \yr{1995}
  \at{Regeneration mechanisms of near-wall turbulence structures}.  \jt{Journal
  of Fluid Mechanics}  \bvol{287},  \pg{317--348}.

\bibitem[Hanratty {\em et~al.\/}(1958)Hanratty, Rosen \&
  Kabel]{hanratty1958effect}
{\sc \au{Hanratty, Thomas~J}, \au{Rosen, Edward~M} \& \au{Kabel, Robert~L}}
  \yr{1958}  \at{Effect of heat transfer on flow field at low reynolds numbers
  in vertical tubes}.  \jt{Industrial \& Engineering Chemistry}  \bvol{50}~(5),
   \pg{815--820}.

\bibitem[He {\em et~al.\/}(2021)He, Tian, Jiang \& He]{he2021turbulence}
{\sc \au{He, J}, \au{Tian, R}, \au{Jiang, PX} \& \au{He, S}} \yr{2021}
  \at{Turbulence in a heated pipe at supercritical pressure}.  \jt{Journal of
  Fluid Mechanics}  \bvol{920},  \pg{A45}.

\bibitem[He {\em et~al.\/}(2016)He, He \& Seddighi]{he2016laminarisation}
{\sc \au{He, S}, \au{He, K} \& \au{Seddighi, M}} \yr{2016}  \at{Laminarisation
  of flow at low reynolds number due to streamwise body force}.  \jt{Journal of
  Fluid mechanics}  \bvol{809},  \pg{31--71}.

\bibitem[Herbert(1983)]{herbert1983secondary}
{\sc \au{Herbert, Thorwald}} \yr{1983}  \at{Secondary instability of plane
  channel flow to subharmonic three-dimensional disturbances}.  \jt{The Physics
  of Fluids}  \bvol{26}~(4),  \pg{871--874}.

\bibitem[Hof {\em et~al.\/}(2010)Hof, De~Lozar, Avila, Tu \&
  Schneider]{hof2010eliminating}
{\sc \au{Hof, Bj{\"o}rn}, \au{De~Lozar, Alberto}, \au{Avila, Marc}, \au{Tu,
  Xiaoyun} \& \au{Schneider, Tobias~M}} \yr{2010}  \at{Eliminating turbulence
  in spatially intermittent flows}.  \jt{science}  \bvol{327}~(5972),
  \pg{1491--1494}.

\bibitem[Hof {\em et~al.\/}(2004)Hof, Van~Doorne, Westerweel, Nieuwstadt,
  Faisst, Eckhardt, Wedin, Kerswell \& Waleffe]{hof2004experimental}
{\sc \au{Hof, Bjorn}, \au{Van~Doorne, Casimir~WH}, \au{Westerweel, Jerry},
  \au{Nieuwstadt, Frans~TM}, \au{Faisst, Holger}, \au{Eckhardt, Bruno},
  \au{Wedin, Hakan}, \au{Kerswell, Richard~R} \& \au{Waleffe, Fabian}}
  \yr{2004}  \at{Experimental observation of nonlinear traveling waves in
  turbulent pipe flow}.  \jt{Science}  \bvol{305}~(5690),  \pg{1594--1598}.

\bibitem[Jackson(2013)]{jackson2013fluid}
{\sc \au{Jackson, JD}} \yr{2013}  \at{Fluid flow and convective heat transfer
  to fluids at supercritical pressure}.  \jt{Nuclear Engineering and Design}
  \bvol{264},  \pg{24--40}.

\bibitem[Jackson \& Li(2002)]{jackson2002influences}
{\sc \au{Jackson, JD} \& \au{Li, J}} \yr{2002}  \at{Influences of buoyancy and
  thermal boundary conditions on heat transfer with naturally-induced flow} .

\bibitem[Kawahara(2005)]{kawahara2005laminarization}
{\sc \au{Kawahara, Genta}} \yr{2005}  \at{Laminarization of minimal plane
  couette flow: going beyond the basin of attraction of turbulence}.
  \jt{Physics of Fluids}  \bvol{17}~(4).

\bibitem[Kawahara \& Kida(2001)]{kawahara2001periodic}
{\sc \au{Kawahara, Genta} \& \au{Kida, Shigeo}} \yr{2001}  \at{Periodic motion
  embedded in plane couette turbulence: regeneration cycle and burst}.
  \jt{Journal of Fluid Mechanics}  \bvol{449},  \pg{291--300}.

\bibitem[Kawahara {\em et~al.\/}(2012)Kawahara, Uhlmann \&
  Van~Veen]{kawahara2012significance}
{\sc \au{Kawahara, Genta}, \au{Uhlmann, Markus} \& \au{Van~Veen, Lennaert}}
  \yr{2012}  \at{The significance of simple invariant solutions in turbulent
  flows}.  \jt{Annual Review of Fluid Mechanics}  \bvol{44},  \pg{203--225}.

\bibitem[Kemeny \& Somers(1962)]{kemeny1962combined}
{\sc \au{Kemeny, GA} \& \au{Somers, EV}} \yr{1962}  \at{Combined free and
  forced-convective flow in vertical circular tubes—experiments with water
  and oil} .

\bibitem[Kerswell \& Tutty(2007)]{kerswell2007recurrence}
{\sc \au{Kerswell, Rich~R} \& \au{Tutty, Owen~R}} \yr{2007}  \at{Recurrence of
  travelling waves in transitional pipe flow}.  \jt{Journal of Fluid Mechanics}
   \bvol{584},  \pg{69--102}.

\bibitem[Khandelwal \& Bera(2015)]{khandelwal2015weakly}
{\sc \au{Khandelwal, Manish~K} \& \au{Bera, P}} \yr{2015}  \at{Weakly nonlinear
  stability analysis of non-isothermal poiseuille flow in a vertical channel}.
  \jt{Physics of Fluids}  \bvol{27}~(6).

\bibitem[Klebanoff {\em et~al.\/}(1962)Klebanoff, Tidstrom \&
  Sargent]{klebanoff1962three}
{\sc \au{Klebanoff, Philip~S}, \au{Tidstrom, KD} \& \au{Sargent, LM}} \yr{1962}
   \at{The three-dimensional nature of boundary-layer instability}.
  \jt{Journal of Fluid Mechanics}  \bvol{12}~(1),  \pg{1--34}.

\bibitem[K{\"u}hnen {\em et~al.\/}(2018)K{\"u}hnen, Song, Scarselli, Budanur,
  Riedl, Willis, Avila \& Hof]{kuhnen2018destabilizing}
{\sc \au{K{\"u}hnen, Jakob}, \au{Song, Baofang}, \au{Scarselli, Davide},
  \au{Budanur, Nazmi~Burak}, \au{Riedl, Michael}, \au{Willis, Ashley~P},
  \au{Avila, Marc} \& \au{Hof, Bj{\"o}rn}} \yr{2018}  \at{Destabilizing
  turbulence in pipe flow}.  \jt{Nature Physics}  \bvol{14}~(4),
  \pg{386--390}.

\bibitem[Lellep {\em et~al.\/}(2022)Lellep, Prexl, Eckhardt \&
  Linkmann]{lellep2022interpreted}
{\sc \au{Lellep, Martin}, \au{Prexl, Jonathan}, \au{Eckhardt, Bruno} \&
  \au{Linkmann, Moritz}} \yr{2022}  \at{Interpreted machine learning in fluid
  dynamics: explaining relaminarisation events in wall-bounded shear flows}.
  \jt{Journal of Fluid Mechanics}  \bvol{942},  \pg{A2}.

\bibitem[Marensi {\em et~al.\/}(2021)Marensi, He \&
  Willis]{marensi2021suppression}
{\sc \au{Marensi, Elena}, \au{He, Shuisheng} \& \au{Willis, Ashley~P}}
  \yr{2021}  \at{Suppression of turbulence and travelling waves in a vertical
  heated pipe}.  \jt{Journal of Fluid Mechanics}  \bvol{919},  \pg{A17}.

\bibitem[Marensi {\em et~al.\/}(2019)Marensi, Willis \&
  Kerswell]{marensi2019stabilisation}
{\sc \au{Marensi, Elena}, \au{Willis, Ashley~P} \& \au{Kerswell, Rich~R}}
  \yr{2019}  \at{Stabilisation and drag reduction of pipe flows by flattening
  the base profile}.  \jt{Journal of Fluid Mechanics}  \bvol{863},
  \pg{850--875}.

\bibitem[McEligot {\em et~al.\/}(1970)McEligot, Coon \&
  Perkins]{mceligot1970relaminarization}
{\sc \au{McEligot, DM}, \au{Coon, CW} \& \au{Perkins, HC}} \yr{1970}
  \at{Relaminarization in tubes}.  \jt{International Journal of Heat and Mass
  Transfer}  \bvol{13}~(2),  \pg{431--433}.

\bibitem[Motoki {\em et~al.\/}(2018)Motoki, Kawahara \&
  Shimizu]{motoki2018optimal}
{\sc \au{Motoki, Shingo}, \au{Kawahara, Genta} \& \au{Shimizu, Masaki}}
  \yr{2018}  \at{Optimal heat transfer enhancement in plane couette flow}.
  \jt{Journal of Fluid Mechanics}  \bvol{835},  \pg{1157--1198}.

\bibitem[Nagata(1990)]{nagata1990three}
{\sc \au{Nagata, Masato}} \yr{1990}  \at{Three-dimensional finite-amplitude
  solutions in plane couette flow: bifurcation from infinity}.  \jt{Journal of
  Fluid Mechanics}  \bvol{217},  \pg{519--527}.

\bibitem[Orr(1907)]{orr1907stability}
{\sc \au{Orr, William~M'F}} \yr{1907} The stability or instability of the
  steady motions of a perfect liquid and of a viscous liquid. part ii: A
  viscous liquid.  \bt{In {\em Proceedings of the Royal Irish Academy. Section
  A: Mathematical and Physical Sciences\/}}, ,  \vol{vol.~27},  \pg{pp.
  69--138}. JSTOR.

\bibitem[Poskas {\em et~al.\/}(2012)Poskas, Poskas \&
  Gediminskas]{poskas2012numerical}
{\sc \au{Poskas, P}, \au{Poskas, R} \& \au{Gediminskas, A}} \yr{2012} Numerical
  investigation of the opposing mixed convection in an inclined flat channel
  using turbulence transition models.  \bt{In {\em Journal of Physics:
  Conference Series\/}}, ,  \vol{vol. 395},  \pg{p. 012098}. IOP Publishing.

\bibitem[Pringle \& Kerswell(2010)]{pringle2010using}
{\sc \au{Pringle, Chris~CT} \& \au{Kerswell, Rich~R}} \yr{2010}  \at{Using
  nonlinear transient growth to construct the minimal seed for shear flow
  turbulence}.  \jt{Physical review letters}  \bvol{105}~(15),  \pg{154502}.

\bibitem[Pringle {\em et~al.\/}(2012)Pringle, Willis \&
  Kerswell]{pringle2012minimal}
{\sc \au{Pringle, Chris~CT}, \au{Willis, Ashley~P} \& \au{Kerswell, Rich~R}}
  \yr{2012}  \at{Minimal seeds for shear flow turbulence: using nonlinear
  transient growth to touch the edge of chaos}.  \jt{Journal of Fluid
  Mechanics}  \bvol{702},  \pg{415--443}.

\bibitem[Reetz {\em et~al.\/}(2019)Reetz, Kreilos \& Schneider]{reetz2019exact}
{\sc \au{Reetz, Florian}, \au{Kreilos, Tobias} \& \au{Schneider, Tobias~M}}
  \yr{2019}  \at{Exact invariant solution reveals the origin of self-organized
  oblique turbulent-laminar stripes}.  \jt{Nature communications}
  \bvol{10}~(1),  \pg{2277}.

\bibitem[Rogers \& Yao(1993)]{rogers1993finite}
{\sc \au{Rogers, BB} \& \au{Yao, LS}} \yr{1993}  \at{Finite-amplitude
  instability of mixed-convection in a heated vertical pipe}.
  \jt{International journal of heat and mass transfer}  \bvol{36}~(9),
  \pg{2305--2315}.

\bibitem[Scheele \& Hanratty(1962)]{scheele1962effect}
{\sc \au{Scheele, George~F} \& \au{Hanratty, Thomas~J}} \yr{1962}  \at{Effect
  of natural convection on stability of flow in a vertical pipe}.  \jt{Journal
  of Fluid Mechanics}  \bvol{14}~(2),  \pg{244--256}.

\bibitem[Scheele {\em et~al.\/}(1960)Scheele, Rosen \&
  Hanratty]{scheele1960effect}
{\sc \au{Scheele, George~F}, \au{Rosen, Edward~M} \& \au{Hanratty, Thomas~J}}
  \yr{1960}  \at{Effect of natural convection on transition to turbulence in
  vertical pipes}.  \jt{The Canadian Journal of Chemical Engineering}
  \bvol{38}~(3),  \pg{67--73}.

\bibitem[Su \& Chung(2000)]{su2000linear}
{\sc \au{Su, Yi-Chung} \& \au{Chung, Jacob~N}} \yr{2000}  \at{Linear stability
  analysis of mixed-convection flow in a vertical pipe}.  \jt{Journal of Fluid
  Mechanics}  \bvol{422},  \pg{141--166}.

\bibitem[Toh \& Itano(2003)]{toh2003periodic}
{\sc \au{Toh, Sadayoshi} \& \au{Itano, Tomoaki}} \yr{2003}  \at{A periodic-like
  solution in channel flow}.  \jt{Journal of Fluid Mechanics}  \bvol{481},
  \pg{67--76}.

\bibitem[Turner \& Turner(1979)]{turner1979buoyancy}
{\sc \au{Turner, John~Stewart} \& \au{Turner, John~Stewart}} \yr{1979} {\em
  Buoyancy effects in fluids\/}.  \publ{Cambridge university press}.

\bibitem[Wang {\em et~al.\/}(2011)Wang, Li, Yu \& Chen]{wang2011investigation}
{\sc \au{Wang, Jianguo}, \au{Li, Huixiong}, \au{Yu, Shuiqing} \& \au{Chen,
  Tingkuan}} \yr{2011}  \at{Investigation on the characteristics and mechanisms
  of unusual heat transfer of supercritical pressure water in vertically-upward
  tubes}.  \jt{International Journal of Heat and Mass Transfer}
  \bvol{54}~(9-10),  \pg{1950--1958}.

\bibitem[Wedin \& Kerswell(2004)]{wedin2004exact}
{\sc \au{Wedin, Hakan} \& \au{Kerswell, Rich~R}} \yr{2004}  \at{Exact coherent
  structures in pipe flow: travelling wave solutions}.  \jt{Journal of Fluid
  Mechanics}  \bvol{508},  \pg{333--371}.

\bibitem[Wibisono {\em et~al.\/}(2015)Wibisono, Addad \&
  Lee]{wibisono2015numerical}
{\sc \au{Wibisono, Andhika~Feri}, \au{Addad, Yacine} \& \au{Lee, Jeong~Ik}}
  \yr{2015}  \at{Numerical investigation on water deteriorated turbulent heat
  transfer regime in vertical upward heated flow in circular tube}.
  \jt{International Journal of Heat and Mass Transfer}  \bvol{83},
  \pg{173--186}.

\bibitem[Willis(2017)]{willis2017openpipeflow}
{\sc \au{Willis, Ashley~P}} \yr{2017}  \at{The openpipeflow navier--stokes
  solver}.  \jt{SoftwareX}  \bvol{6},  \pg{124--127}.

\bibitem[Willis(2019)]{willis2019equilibria}
{\sc \au{Willis, Ashley~P}} \yr{2019}  \at{Equilibria, periodic orbits and
  computing them}.  \jt{arXiv preprint arXiv:1908.06730} .

\bibitem[Yao(1987)]{yao1987fully}
{\sc \au{Yao, LS}} \yr{1987}  \at{Is a fully-developed and non-isothermal flow
  possible in a vertical pipe?}  \jt{International journal of heat and mass
  transfer}  \bvol{30}~(4),  \pg{707--716}.

\bibitem[Yoo(2013)]{yoo2013turbulent}
{\sc \au{Yoo, Jung~Yul}} \yr{2013}  \at{The turbulent flows of supercritical
  fluids with heat transfer}.  \jt{Annual review of fluid mechanics}
  \bvol{45},  \pg{495--525}.

\bibitem[You {\em et~al.\/}(2003)You, Yoo \& Choi]{you2003direct}
{\sc \au{You, Jongwoo}, \au{Yoo, Jung~Y} \& \au{Choi, Haecheon}} \yr{2003}
  \at{Direct numerical simulation of heated vertical air flows in fully
  developed turbulent mixed convection}.  \jt{International Journal of Heat and
  Mass Transfer}  \bvol{46}~(9),  \pg{1613--1627}.

\bibitem[Zhang {\em et~al.\/}(2020)Zhang, Xu, Liu \& Dang]{zhang2020review}
{\sc \au{Zhang, Shijie}, \au{Xu, Xiaoxiao}, \au{Liu, Chao} \& \au{Dang,
  Chaobin}} \yr{2020}  \at{A review on application and heat transfer
  enhancement of supercritical co2 in low-grade heat conversion}.  \jt{Applied
  Energy}  \bvol{269},  \pg{114962}.

\bibitem[Zhao {\em et~al.\/}(2018)Zhao, Zhu, Ge, Liu \& Li]{zhao2018direct}
{\sc \au{Zhao, Pinghui}, \au{Zhu, Jiayin}, \au{Ge, Zhihao}, \au{Liu, Jiaming}
  \& \au{Li, Yuanjie}} \yr{2018}  \at{Direct numerical simulation of turbulent
  mixed convection of lbe in heated upward pipe flows}.  \jt{International
  Journal of Heat and Mass Transfer}  \bvol{126},  \pg{1275--1288}.

\end{thebibliography}


\end{document}